\documentclass[aps,prx,reprint,superscriptaddress]{revtex4-2}
\usepackage[]{graphicx}
\usepackage{amsmath,amssymb}
\usepackage{newtxtext}
\usepackage{newtxmath}
\usepackage{nicefrac}
\usepackage[utf8]{inputenc}
\usepackage[nodayofweek,ddmmyy]{datetime}
\usepackage[x11names]{xcolor}
\usepackage[unicode,colorlinks,citecolor=Blue3,linkcolor=Blue3,bookmarks=true]{hyperref}
\usepackage[normalem]{ulem}
\usepackage{transparent}

\newcommand{\sign}{\mathop{\rm sign}\nolimits}

\renewcommand{\le}{\leqslant}

\renewcommand{\Im}{\mathop{\rm Im}\nolimits}

\newcommand{\fulld}[2]{\dfrac{d#1}{d#2}}
\newcommand{\fulldd}[2]{\dfrac{d^2#1}{d#2^2}}

\newcommand{\svector}[2]{\begin{pmatrix}#1 \\ #2 \end{pmatrix}}

\newcommand{\smatrix}[4]{\begin{pmatrix}#1 & #2 \\ #3 & #4\end{pmatrix}}

\begin{document}

\title{Trajectories without quantum uncertainties in composite systems with disparate energy spectra}

\author{Emil Zeuthen}
\email{zeuthen@nbi.ku.dk}
\affiliation{Niels Bohr Institute, University of Copenhagen, DK-2100 Copenhagen, Denmark}

\author{Eugene S.\ Polzik}
\affiliation{Niels Bohr Institute, University of Copenhagen, DK-2100 Copenhagen, Denmark}

\author{Farid Ya.\ Khalili}
\email{farit.khalili@gmail.com}
\affiliation{Russian Quantum Center, Skolkovo IC, Bolshoy Bulvar 30, bld.\ 1, Moscow, 121205, Russia}
\affiliation{NTI Center for Quantum Communications, National University of Science and Technology MISiS, Leninsky prospekt 4, Moscow 119049, Russia}

\begin{abstract}

It is well established that measurement-induced quantum back action (QBA) can be eliminated in composite systems by engineering so-called quantum-mechanics-free subspaces (QMFSs) of commuting variables, leading to a trajectory of a quantum system without quantum uncertainties. 
This situation can be realized in a composite system that includes a negative-mass subsystem, which can be implemented by, e.g., a polarized spin ensemble or a two-tone-driven optomechanical system. 
The realization of a trajectory without quantum uncertainties implies entanglement between the subsystems, and allows for measurements of motion, fields and forces with, in principle, unlimited precision. To date, these principles have been developed theoretically and demonstrated experimentally for a number of composite systems. However, the utility of the concept has been limited by the dominating requirement of close proximity of the resonance frequencies of the system of interest and the negative-mass reference system, and by the need to embed the subsystems in a narrowband cavity, which could be problematic while at the same time achieving good overcoupling. Here we propose a general approach which overcomes these limitations by employing periodic modulation of the driving fields (e.g., two-tone driving) in combination with coherent or measurement-based anti-noise paths. This approach makes it possible to engineer a QMFS of two systems with vastly different spectra and with arbitrary signs of their masses, while dispensing with the need to embed the subsystems in a sideband-resolving cavity. We discuss the advantages of this novel approach for applications such as QBA evasion in gravitational wave detection, force sensing, and entanglement generation between disparate systems.

\end{abstract}

\maketitle

\section{Introduction}\label{sec:intro}

The existence of non-commuting observables in quantum mechanics directly implies the concept of measurement-induced quantum back action (QBA)~\cite{Neumann_e}. In the case of Gaussian quantum states and measurements, which is the most relevant one for the macroscopic quantum measurements that we consider here, the impact of QBA is bounded by the Schr\"odinger-Robertson uncertainty relation~\cite{Schroedinger_PPAS_19_296_1930}.
The successful evasion of QBA, by channeling it into unobserved degrees of freedom, is a central ingredient in a number of quantum protocols, e.g., entanglement generation, quantum teleportation, and quantum sensing.

In particular, in the linear force and displacement measurement schemes in which the QBA is uncorrelated with the measurement imprecision noise, the sensitivity is constrained by the Standard Quantum Limit (SQL), at which these two contributions balance~\cite{67a1eBr, Caves_RMP_52_341_1980, 80a1BrThVo}. Note that the sensitivity of the modern laser gravitational-wave detectors (GWDs) is approaching the SQL, and the means of overcoming it are actively discussed in the literature, see, e.g., the review articles~\cite{12a1DaKh, 19a1DaKhMi}.

In its conceptually simplest form, QBA evasion can be achieved by measuring a quantum non-demolition (QND) variable of the probe object, that is, one that is autocommuting at different moments of time --- for example one of two quadratures of a harmonic oscillator~\cite{78a1eBrKhVo, Thorne1978, Caves_RMP_52_341_1980} or the momentum of a free mass~\cite{90a1BrKh}. The QBA in this case is channeled into the canonically conjugate observable (the second quadrature or the position, respectively).

A more general approach to QND measurements, involving more than a single variable, has emerged, namely a measurement with respect to a designed reference frame characterized by an {\it effective negative mass}. In the case where this reference system is a harmonic oscillator, this amounts to a negative eigenfrequency. An example of such an oscillator is a collective spin of an atomic ensemble in a magnetic field prepared in an energetically maximal state so that a spin flip \emph{reduces} the energy of the ensemble. This idea had been implicitly utilized for the first time for the experimental demonstration of entanglement between two collective spin ensembles, a positive- and a negative-frequency one~\cite{Julsgaard_Nature_413_400_2001}. Extending this idea to a hybrid setting, several proposals have considered the combination of a negative-frequency spin oscillator with a positive-frequency mechanical oscillator~\cite{Hammerer_PRL_102_020501_2009, Bariani_PRA_92_043817_2015, Motazedifard_NJP_18_073040_2016, 18a1KhPo} (the term ``negative-mass oscillator'' was coined in Ref.~\cite{Hammerer_PRL_102_020501_2009}). Besides using the negative-frequency spin ensembles, it has been proposed to implement a negative-mass reference frame using an optical~\cite{Tsang_PRL_105_123601_2010} or a mechanical~\cite{Woolley_PRA_87_063846_2013} system, as well as a Bose-Einstein condensate~\cite{Zhang_PRA_88_043632_2013}.  

Measurement in a negative-mass reference frame has been used to demonstrate QBA-free magnetic-field sensing~\cite{Wasilewski_PRL_104_133601_2010}, and has subsequently been referred to as {\it trajectories without quantum uncertainties}~\cite{Polzik_AnnPhys_527_A15_2014}. 
Recently, experimental demonstrations of the concept of the negative-mass reference frame have been carried out with a purely optical system~\cite{Zander2021}, a hybrid system of distant mechanical and spin oscillators~\cite{Moeller_Nature_547_191_2017, Thomas_NPhys_17_228_2021}, and with two mechanical oscillators coupled to a common superconducting microwave cavity~\cite{Ockeloen-Korppi_PRL_117_140401_2016,Lepinay_Science_372_625_2021}.

Formally, measurements performed in a negative-mass (or \nobreakdash-frequency) reference frame can be described in terms of a set of QBA-free, commuting variables decoupled from the variables subject to QBA, hence constituting a so-called {\it quantum-mechanics-free subspace} (QMFS)~\cite{Tsang_PRX_2_031016_2012}. Because two (or more) degrees of freedom are involved in this case, in contrast to an orthodox QND measurement, the QMFS can be used not only for sensing, but also for quantum entanglement and teleportation applications.

The bulk of work within this emerging area, and all experimental demonstrations, have so far been dealing with subsystems characterized by nearly identical spectra. In this case, the susceptibilities of the subsystems should be identical (up to a frequency-independent numerical factor) and have opposite signs, which ensure the QMFS. For far-off-resonant sensing, the resonance frequencies of the subsystems may differ as long as the band of signal frequencies is far-off-resonant with respect to both resonances~\cite{18a1KhPo}. In force sensing applications, the effective resonance frequencies can be manipulated by means of the virtual rigidity effect~\cite{19a1ZePoKh}. Nonetheless, even in these cases the bare resonance frequencies of the subsystems have to be of roughly the same order of magnitude.

In principle, this limitation can be lifted by means of multi-carrier (modulated) drive fields. However, the modulated coupling will generally induce QBA from unwanted frequency components (see, e.g., Ref.~\cite{Buchmann_PRA_92_013851_2015}) that must be suppressed in some way. In existing schemes based on a two-tone drive, this issue is addressed by embedding the system in an electromagnetic cavity with one or more sideband-resolving modes~\cite{Caves_RMP_52_341_1980, Woolley_PRA_87_063846_2013, Tan_PRA_87_022318_2013}. This strategy imposes a design constraint on hybrid systems which (in many cases) will be an impediment to achieving the large cavity overcoupling required for efficient quantum linking of distant systems via a traveling light field.

Here we propose a universal approach, using periodically modulated drive fields, that enables the realization of a QMFS between two arbitrary oscillators having, in principle, completely different physical nature, arbitrary resonance frequencies (including the free mass case), and arbitrary signs of the effective masses. Crucially, we propose methods for suppression of the aforementioned unwanted QBA frequency components which can be implemented in the bad-cavity limit, thus eliminating the technical challenge of embedding systems in narrowband cavities.

The flexibility in the choice of both optical carrier frequencies and the subsystems’ resonance frequencies could be of special interest for GWDs, which use very low-frequency ($\sim1\,{\rm Hz}$) pendulums as probe objects. The methods for broadband suppression of QBA in GWDs which are considered the most probable candidates for implementation, require either the use of additional expensive kilometer-scale filter cavities to create the frequency-dependent cross-correlations between the imprecision noise and the back action noise~\cite{Unruh1982, 02a1KiLeMaThVy} or radical modification of the interferometer to implement the quantum speedmeter topology~\cite{90a1BrKh, 00a1BrGoKhTh, Chen2002}. In this respect, the prospect of using instead a table-top system, based either on a small-scale optomechanical setup or an atomic spin ensemble as the negative-mass reference frame of the QMFS, could be very attractive.

The paper is organized as follows. We review the principles of the QMFS in Subsec.~\ref{sec:intro-QMFS} and the basic topologies for implementing them in Subsec.~\ref{sec:topologies}. In Subsec.~\ref{subsec:strob_intro} we introduce semi-qualitatively the principle of a QMFS with disparate frequency scales.
Section~\ref{sec:univ-downconv} contains the detailed description and analysis of our frequency-conversion scheme for realizing a QMFS under such circumstances, including methods for suppression of the extraneous (high-frequency) QBA components.
In Section~\ref{sec:applications}, we consider the application of our scheme to QBA-evading pulsed and continuous force sensing, including sensitivity estimates for laser GWDs, and entanglement generation between distant systems.
Finally, in Section~\ref{sec:Conclusion-outlook}, we recapitulate and outline the future prospects for our work.

\section{Quantum-mechanics-free subspaces and periodic coupling envelopes}\label{sec:QMFS}

\subsection{Introduction to quantum-mechanics-free subspaces}\label{sec:intro-QMFS}

Consider an oscillator with the Hamiltonian $\hat{H}=\hat{H}_0 + \hat{H}^\prime$, where
\begin{equation}\label{eq:H-osc}
  \hat{H}_0=\frac{\Omega_0}{2}\biggl(\rho\hat{x}^2 + \frac{\hat{p}^2}{\rho}\biggr)
\end{equation}
is the free Hamiltonian, $\hat{H}^\prime$ describes all couplings of the oscillator with other degrees of freedom (coherent probing, thermal reservoir, etc.),
\begin{equation}
  \rho = m\Omega_0
\end{equation}
is the characteristic impedance of the oscillator, which we assume to be positive, and $\Omega_0$ is the evolution frequency, which could be both positive or negative. From this it follows that $\sign\Omega_0=\sign m$ and, hence, that a negative evolution frequency $\Omega_0<0$ for an oscillator is equivalent to it having a negative mass $m<0$. While we have referred to the concept of a negative mass in the Introduction, we will henceforth use the (equivalent) term ``negative (evolution) frequency'', as it relates more closely to the mathematical formulation employed here.

We introduce the dimensionless oscillator position $\hat{X}$ and momentum $\hat{P}$ via
\begin{equation}\label{norm_XP}
  \hat{X} = \frac{\hat{x}}{\sqrt{\hbar/\rho}} \,, \quad
  \hat{P} = \frac{\hat{p}}{\sqrt{\hbar\rho}} \,,
\end{equation}
satisfying the commutation relation
\begin{equation}\label{ur_xp}
  [\hat{X}, \hat{P}] = i \,.
\end{equation}
In this notation the free Hamiltonian~\eqref{eq:H-osc} reads
\begin{equation}\label{eq:H-osc_norm}
  \hat{H}_0=\frac{\hbar\Omega_0}{2}(\hat{X}^2 + \hat{P}^2)  \,.
\end{equation}

For the typical scenario of a weak, continuous measurement of the oscillator, it is convenient to consider its
slowly-varying quadrature operators $\hat{\mathcal{X}}$ and $\hat{\mathcal{P}}$ which are related to the original variables by a rotation with the angular frequency $\Omega_0$ in the $(\hat{X}, \hat{P})$-plane (the rotating-frame picture):
\begin{equation}\label{eq:slowly-rot-vars}
  \svector{\hat{X}(t)}{\hat{P}(t)}
  = \smatrix{\cos \Omega_0t}{\sin \Omega_0t}{-\sin \Omega_0t}{\cos \Omega_0t}
      \svector{\hat{\mathcal{X}}(t)}{\hat{\mathcal{P}}(t)} .
\end{equation}
Assuming that $\hat{\mathcal{X}}$ and $\hat{\mathcal{P}}$ evolve slowly compared to the time-scale defined by the oscillation period $2\pi/|\Omega_0|$, a continuous measurement of, e.g., $\hat{X}$ over several periods constitutes a simultaneous measurement of the two non-commuting quadratures $\hat{\mathcal{X}}$ and $\hat{\mathcal{P}}$. The precision of this measurement is constrained by the Heisenberg uncertainty relation since their commutation relation has the same form as Eq.~\eqref{ur_xp}:
\begin{equation}
  [\hat{\mathcal{X}}, \hat{\mathcal{P}}] = i \,.
\end{equation}

Consider now two such oscillators with evolution frequencies $\Omega_{1,2}$ obeying
\begin{equation}\label{eq:counter-rot-cond}
\Omega_1=-\Omega_2\equiv\Omega_0,
\end{equation}
a so-called counter-rotating pair of oscillators. Assume now that we continuously measure one of the joint variables $\hat{X}_1\pm\hat{X}_2$ or $\hat{P}_1\pm\hat{P}_2$, e.g.,  $\hat{X}_1+\hat{X}_2$. We observe that Eqs.~(\ref{eq:slowly-rot-vars}, \ref{eq:counter-rot-cond}) imply
\begin{equation}\label{eq:EPR-osc}
\hat{X}_1(t)+\hat{X}_2(t) = [\hat{\mathcal{X}}_1(t)+\hat{\mathcal{X}}_2(t)]\cos\Omega_0 t
  + [\hat{\mathcal{P}}_1(t)-\hat{\mathcal{P}}_2(t)]\sin\Omega_0 t \,,
\end{equation}
and hence a continuous measurement of $\hat{X}_1+\hat{X}_2$ amounts to a simultaneous measurement of the \emph{commuting} pair of Einstein-Podolsky-Rosen (EPR) variables:
\begin{equation}\label{eq:EPR-comm}
  [\hat{\mathcal{X}}_1+\hat{\mathcal{X}}_2,\,\hat{\mathcal{P}}_1-\hat{\mathcal{P}}_2] = 0 \,.
\end{equation}
The sensitivity of this measurement is not constrained by the Heisenberg uncertainty relation. Therefore, these EPR variables of the two counter-rotating oscillators constitute a QMFS. The crucial relative minus sign between $\hat{\mathcal{P}}_1$ and $\hat{\mathcal{P}}_2$ in Eqs.~\eqref{eq:EPR-osc} and~\eqref{eq:EPR-comm} arises from the requirement~\eqref{eq:counter-rot-cond}, which thus plays a key role in our approach to QMFSs. A central element of the present work is a frequency-conversion technique for fulfilling Eq.~\eqref{eq:counter-rot-cond} for two arbitrary and potentially disparate oscillators.

The preceding example, described by Eqs.~(\ref{eq:counter-rot-cond}-\ref{eq:EPR-comm}), conveys the essence of the formal, necessary conditions for establishing a QMFS in the approach taken here. These can be summarized as the following three central requirements:
\begin{enumerate}
\item[(a)] the engineering of a joint quantum measurement on separated systems that could be probed by disparate optical carrier frequencies $\omega_{o,j}$;
\item[(b)] matching of the subsystems' (absolute) evolution frequencies $|\Omega_1|=|\Omega_2|$ that determine the frequency scale of their QBA response relative to the optical carrier; and
\item[(c)] the realization of counter-rotating oscillators: $\sign\Omega_1=-\sign\Omega_2$.
\end{enumerate}
The following two subsections give a qualitative discussion of these three requirements and an overview of how the present work proposes to fulfill them.

\subsection{Topologies for QBA evasion}\label{sec:topologies}

\begin{figure}[htbp]
\centering
\def\svgwidth{\columnwidth}
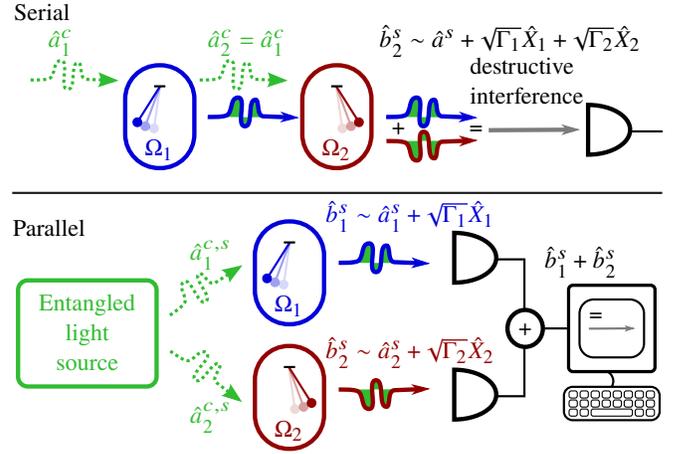
\caption{Basic single-pass topologies for quantum noise evasion. In the serial topology, the (stochastic) QBA forces $\propto\hat{a}^c_j$ (green, dashed arrows) on the two oscillators (pendula) are perfectly correlated $\hat{a}^c_2=\hat{a}^c_1$ (in the absence of optical losses) simply due to the fact that the same itinerant field is probing the two subsystems. The QBA responses (thick arrows with green filling) of both oscillators are mapped into the phase quadrature $\hat{b}^s_2$ of the output light. These contributions will interfere destructively assuming the counter-rotating frequency configuration $\Omega_1=-\Omega_2$ with matched light-oscillator coupling rates $\Gamma_1=\Gamma_2$. In the parallel topology, a source of entangled light correlates the QBA forces $\text{Cov}[\hat{a}^c_1,\hat{a}^c_2]>0$ of the two meter fields and anti-correlates the measurement imprecision noises $\text{Cov}[\hat{a}^s_1,\hat{a}^s_2]<0$ so that both interfere destructively when the two output photocurrents are combined in post-processing, again provided that $\Omega_1=-\Omega_2$ and $\Gamma_1=\Gamma_2$.}
\label{fig:topologies}
\end{figure}

Our qualitative overview starts with addressing requirement (a) by reviewing topologies for engineering a joint, QBA-evading measurement, such as the one implied by Eq.~\eqref{eq:EPR-osc}.
We do so assuming that requirements (b) and (c) are met to begin with. Afterwards, in Subsec.~\ref{subsec:strob_intro}, we give a preview of how to achieve (b) and (c) using our down-conversion technique, before rigorously introducing it in Sec.~\ref{sec:univ-downconv}.

Consider two oscillators $j\in\{1,2\}$ that we wish to subject to a joint measurement. A joint measurement on a number of subsystems (two in this case) can, for our purposes, be characterized as a measurement for which the meter imprecision and QBA noise sources are correlated across the subsystems. The most straightforward way to implement this requirement is to probe the oscillators in cascade by the same light beam, as shown in Fig.~\ref{fig:topologies}(top)~\cite{Julsgaard_Nature_413_400_2001, Hammerer_PRL_102_020501_2009, Moeller_Nature_547_191_2017}.

At the same time, this topology enforces the use of the same light carrier frequency $\omega_{o,j}=\omega_o$ in both subsystems, which could be impossible for systems with different spectra (atoms, molecules, etc) or could be problematic due to technological limitations. In those cases, one could use the \emph{parallel} topology shown in Fig.~\ref{fig:topologies}(bottom), which can accommodate unequal carrier frequencies $\omega_{o,1}\neq \omega_{o,2}$~\cite{18a1KhPo}. This topology (first proposed in Ref.~\cite{Ma_NPhys_13_776_2017} in a different context) relies on two optical fields prepared in an entangled quantum state generated, for example, by a non-degenerate optical parametric oscillator. In the parallel case, the degree of correlation between the two optical fields is set by the degree of two-mode squeezing, for which large values have been demonstrated~(see, e.g., Refs.~\cite{Wang_AppSci_8_330_2018,Brasil_arXiv_2021}). Regardless of the finite degree of correlation in the parallel topology, it shares the crucial trait with the serial topology that the QBAs on the two subsystems are correlated. 

In Ref.~\cite{Karg_PRA_99_063829_2019}, the generalized serial topology, which could involve an arbitrary number of oscillators and loops of the light field, was presented (but no considerations regarding frequency conversion were given). 
An instance of such a topology has been demonstrated experimentally outside the QBA-dominated regime~\cite{Karg_Science_369_174_2020}.
However, here we limit ourselves to the aforementioned (single-pass) serial and parallel scenarios which are the most practical candidates for hybrid and/or distributed systems.

\subsection{Linking disparate frequency scales of the subsystems}\label{subsec:strob_intro}

We now turn to the requirements (b) and (c) [equivalent to Eq.~\eqref{eq:counter-rot-cond}], which are generally not fulfilled in realistic hybrid systems. In order to satisfy them, we propose the use of modulated probing, which is a well-known technique in quantum measurement theory. In particular, the first proposals for QBA-evading schemes relied on the probing strength varying with a periodicity $\tilde{T}=2\pi/|\Omega_0|$ matching that of the oscillator evolution (at frequency $\Omega_0$), in order to measure a \emph{single} oscillator quadrature: the stroboscopic measurement scheme of Ref.~\cite{78a1eBrKhVo}, the mechanical coordinate and momentum sensing scheme of Ref.~\cite{Thorne1978}, and the two-tone drive scheme of Ref.~\cite{Caves_RMP_52_341_1980}.
We will refer to this situation as \emph{resonant} periodic probing.
In contrast, our scheme makes use of \emph{detuned} periodic probing, where $\tilde{T}\neq 2\pi/|\Omega_0|$,
resulting in a measurement of \emph{both} quadratures of an effective oscillator with a new, effective resonance frequency, which can be either positive or negative.
Hence this constitutes a frequency-conversion mechanism. The simultaneous measurement of both quadratures (as opposed to only a single one) is essential to constructing non-trivial QMFSs of commuting variables~\eqref{eq:EPR-comm}. More concretely, in terms of applications, a measurement of both oscillator quadratures is required for, e.g., EPR entanglement of oscillators and the simultaneous detection of \emph{both} phases of a classical signal force.

Now we describe semi-qualitatively the principles behind our frequency-conversion technique in preparation for its rigorous derivation in Sec.~\ref{sec:univ-downconv}. We begin by noting that if an oscillator is observed not continuously, but only stroboscopically at the moments of time $t_n = \pi n/\tilde{\Omega}$ where the probing frequency $\tilde{\Omega}>0$ is slightly detuned from the bare oscillator frequency by the amount
  \begin{equation}\label{Omega_p}
    \Lambda \equiv |\Omega_0| - \tilde{\Omega} \,, \quad |\Lambda| \ll |\Omega_0| \,,
  \end{equation}
then the oscillator appears to the observer as one with effective eigenfrequency (which could be either positive or negative)
\begin{equation}\label{Omega_eff}
  \Omega_{\rm eff} = s\Lambda \,,
\end{equation}
where the factor
\begin{equation}
  s = \sign\Omega_0
\end{equation}
takes into account that the eigenfrequency $\Omega_0$ could be negative; note that Eq.~\eqref{Omega_p} implies $|\Omega_{\text{eff}}|\ll|\Omega_0|$ whereby only down-conversion is permitted. This effect is seen by evaluating Eq.~\eqref{eq:slowly-rot-vars} at $t=t_n$ under the assumption~\eqref{Omega_p},
\begin{equation}\label{eq:stroboscopic-rot}
\left(\begin{array}{c}
\hat{X}(t_n)\\
\hat{P}(t_n)
\end{array}\right)\approx\left(\begin{array}{cc}
\cos \Omega_{\text{eff}} t_n & \sin \Omega_{\text{eff}} t_n \\
-\sin \Omega_{\text{eff}} t_n & \cos \Omega_{\text{eff}} t_n
\end{array}\right)\left(\begin{array}{c}
\hat{\mathcal{X}}(t_n)\\
\hat{\mathcal{P}}(t_n)
\end{array}\right) .
\end{equation}
In Eq.\,\eqref{eq:stroboscopic-rot} we have removed a common prefactor $(-1)^n$ as can be done in the post-processing of the measurement record for $\hat{X}(t_n)$ and/or $\hat{P}(t_n)$. Note that the slowly-varying quadratures $(\hat{\mathcal{X}},\hat{\mathcal{P}})$ are defined with respect to $\Omega_0$, see Eq.~\eqref{eq:slowly-rot-vars}.

The same outcome can be achieved with a two-tone probe field, i.e., with a harmonically varying coupling strength $\propto\cos\tilde{\Omega}t$ between the oscillator and the traveling light field. Averaging out rapid dynamics at frequencies $\sim2\tilde{\Omega}$ (as can be done, e.g., in post-processing) we find for the time-modulated readout of, e.g., $\hat{X}$ that the oscillator signal is
\begin{equation}\label{eq:two-tone-rot}
  \overline{\hat{X}(t)\cos\tilde{\Omega}t}
  \approx \hat{\mathcal{X}}(t)\cos\Omega_{\text{eff}} t + \hat{\mathcal{P}}(t)\sin\Omega_{\text{eff}} t\,,
\end{equation}
which is analogous to the stroboscopic readout of $\hat{X}(t_n)$ in Eq.~\eqref{eq:stroboscopic-rot} (see the rigorous analysis in Sec.~\ref{sec:univ-downconv}).

Therefore, using periodic, non-stationary measurements in one of the two subsystems that we wish to match, it is possible to change the sign of its ``perceived'' resonance frequency and shift its absolute value to match the resonance frequency of the second subsystem, constructing thus a QMFS for subsystems with very different resonance frequencies.
Alternatively, the technique can be applied to both systems in a manner so that $s_1\Lambda_1=-s_2\Lambda_2 \Leftrightarrow\Omega_{1,\text{eff}}=-\Omega_{2,\text{eff}}$, where $s_i\equiv\sign\Omega_i$.

However, in the context of quantum measurements, equations~\eqref{eq:stroboscopic-rot} and~\eqref{eq:two-tone-rot} only tell half of the story; one must bear in mind that the modulation envelope (in the examples given here: a train of stroboscopic pulses or a harmonic function) also determines the spectral composition of the associated QBA. This aspect is a key consideration in establishing a QMFS since the latter relies on the engineering of appropriate quantum noise correlations between the subsystems involved.

As a qualitative prelude to Sec.~\ref{sec:univ-downconv}, we now sketch how our periodic driving schemes can accomplish the desired matching of the sideband frequency scales. Let us consider the scenario of two systems with disparate bare resonance frequencies $|\Omega_1|\ll|\Omega_2|$, for which our scheme is most pertinent. This implies that down-conversion must be applied to oscillator 2 in order to engineer the matching of (effective) resonance frequencies $\Omega_{1}=-\Omega_{2,\text{eff}}$. As we will see, we can accomplish this provided that the following conditions are met:
\begin{equation}\label{RWA2}
|\Omega|,|\Lambda|,\gamma_{1,2} \ll \tilde{\Omega}\,.
\end{equation}
Here $\gamma_{1,2}$ are the half-bandwidths of the two oscillator systems, and the assumption on the Fourier frequency $\Omega$ reflects the range of frequencies we are interested in.

\begin{figure}[htb]
\centering
\def\svgwidth{1\columnwidth}
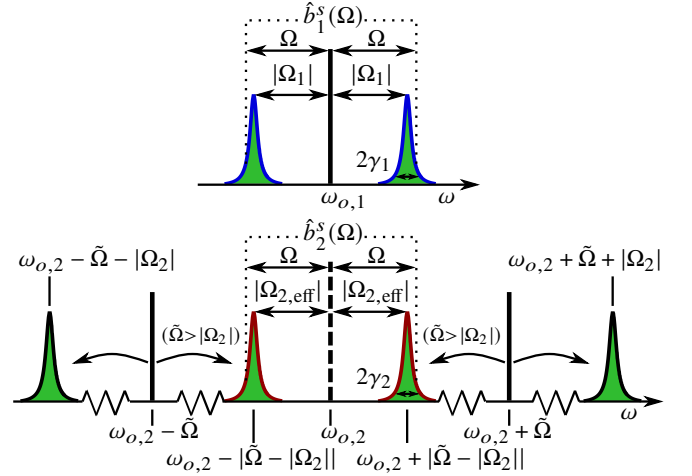
\caption{Basic principle behind the universal frequency conversion technique as exemplified by two-tone probing (lower part) and juxtaposed with the case of conventional single-tone driving (upper part). These schemes for coupling an oscillator to a traveling light field are illustrated by their individual output spectra; the two-photon (homodyne) phase quadrature $\hat{b}_j^s(\Omega)$ results from folding the spectrum around the local oscillator frequency $\omega_{\text{LO},j}=\omega_{o,j}$. (top) Single-tone coupling of an oscillator with eigenfrequency $\Omega_1$ using a tone at $\omega_{o,1}$ (solid line); the sidebands peak at frequencies $\approx\omega_{o,1}\pm\Omega_1$ (true in the high-$Q$-limit, $|\Omega_j|/(2\gamma_j)\gg1$; sideband widths are exaggerated in the figure for illustrative purposes). (bottom) Two-tone coupling scheme using coherent tones at $\omega_{o,2}\pm\tilde{\Omega}$ (solid lines) with mean frequency $\omega_{o,2}$ (dashed line) resulting in the effective oscillator evolution frequency $\Omega_{2,\text{eff}}\equiv\Lambda\sign\Omega_2=(|\Omega_2|-\tilde{\Omega})\sign\Omega_2$.  The curved arrows indicate the sidebands generated by the respective drive tones in the particularly interesting case $\tilde{\Omega}>|\Omega_2|\Leftrightarrow\Lambda<0$, where the sign of the evolution frequency is flipped in the downconversion process $\sign\Omega_{2,\text{eff}}=-\sign\Omega_2$. This reflects the circumstance that for $\tilde{\Omega}>|\Omega_2|$ the upper (lower) central sideband arises as the lower (upper) sideband of the originating drive tone. If a joint measurement is performed on oscillators 1 and 2, broadband destructive interference between the QBA responses contained in the green, central sidebands is possible if $\Omega_1=-\Omega_{2,\text{eff}}$.}
\label{fig:carrier-sidebands}
\end{figure}

We focus on the example of (detuned) two-tone driving for performing the down-conversion, see Fig.~\ref{fig:carrier-sidebands}. Here, the $\Omega_1$ system is coupled to a single-tone driving field $\omega_{o,1}$ as shown in the upper panel, leading to the two sideband frequencies $\omega_{o,1}\pm\Omega_1$. The sideband at $\omega_{o,1}+\Omega_1$ is generated by light-oscillator interaction of the beamsplitter (BS) type, whereas that at $\omega_{o,1}-\Omega_1$ is generated by parametric down-conversion (PDC). Hence, the sign of $\Omega_1$ determines which interaction type is responsible for the lower- and higher-frequency (Stokes and anti-Stokes) sidebands, respectively.

The $\Omega_2$ system is coupled to a driving field consisting of two coherent tones with frequencies $\omega_{o,2}\pm\tilde{\Omega}$ leading to the four sideband frequencies $\omega_{o,2}\pm\tilde{\Omega} \pm^\prime \Omega_2$ as shown in the lower panel. Let $\tilde{\Omega}=\Omega_1\sign\Omega_2+|\Omega_2|$, which is equivalent to $\Lambda=-\Omega_1\sign\Omega_2$. In this case, $\Omega_{2,\text{eff}} = \Lambda\sign\Omega_2= -\Omega_1$ [see Eq.~\eqref{Omega_eff}], i.e., the  condition~\eqref{eq:counter-rot-cond} is satisfied (as assumed in Fig.~\ref{fig:carrier-sidebands}).

The physics of the sign of $\Omega_{2,\text{eff}}$ can be understood from the spectral picture in Fig.~\ref{fig:carrier-sidebands} as follows.
The two coherent tones generate four sidebands: two lower-frequency (Stokes) ones centered at $\omega_{o,2}\pm\tilde{\Omega} - |\Omega_2|$ and two higher-frequency (anti-Stokes) ones at $\omega_{o,2}\pm\tilde{\Omega} + |\Omega_2|$. Out of these, the two closest to $\omega_{o,2}$, i.e., at $\omega_{o,2}\pm\Lambda$, are the ones that constitute the {\it effective} Stokes/anti-Stokes pair of the effective oscillator with eigenfrequency $\Omega_{2,\text{eff}}=\Lambda \sign\Omega_2$.

In particular, if $\Lambda>0$, then the high-frequency coherent tone $\omega_{o,2}+\tilde{\Omega}$ creates the effective Stokes sideband at $\omega_{o,2}-|\Lambda|$, whereas if $\Lambda<0$, it creates the effective anti-Stokes one at $\omega_{o,2}+|\Lambda|$. In both cases, if $\sign\Omega_2>0$, then PDC-type interaction is involved, whereas if $\sign\Omega_2<0$, it is the BS type. For the low-frequency tone $\omega_{o,2}-\tilde{\Omega}$, analogous statements hold where all inequalities are inverted.

In all cases, if $\Lambda\sign\Omega_2>0$, then, as usual, the effective Stokes and anti-Stokes sidebands are created by PDC and BS interactions, respectively. But if $\Lambda\sign\Omega_2<0$, then we achieve the inverted scenario of the effective Stokes sideband generated by a BS interaction, and the effective anti-Stokes one -- by a PDC interaction. This gives rise to the negative-mass character of the effective oscillator.

Returning now to the joint measurement of the oscillators in the configuration $\Omega_{2,\text{eff}}=-\Omega_1$, the central sidebands at $\omega_{o,2}\pm\Omega_{2,\text{eff}}$ cancel their counterparts at $\omega_{o,1}\mp\Omega_1$ when combined in the joint measurement. 
The extraneous sidebands at $\omega_{o,2}\pm(\tilde{\Omega}+|\Omega_2|)$ (furthest away from $\omega_{o,2}$) can be removed in post-processing, however the associated extraneous QBA (not shown in Fig.~\ref{fig:carrier-sidebands}) must be cancelled for efficient operation -- this matter is a central topic of this work.

\section{Universal down-conversion of an oscillator}\label{sec:univ-downconv}

\subsection{Oscillator with modulated coupling to itinerant light field}\label{subsec:osc-modulated-coup}

Now we turn to the rigorous description of our proposal for engineering the effective evolution frequency of an oscillator, thus providing a means for realizing the counter-rotating condition~\eqref{eq:counter-rot-cond} across a range of hybrid systems.
The central idea of this frequency-downconversion scheme is the realization that an oscillator with resonance frequency $\Omega_0$ subject to a suitable periodically modulated drive field, can be made to act like an oscillator of arbitrary effective resonance frequency, as discussed qualitatively in Subsec.~\ref{subsec:strob_intro}.

Consider an oscillator parametrically coupled to a traveling optical field with a time-dependent coupling strength,
\begin{equation}
  E(t) = E_0k(t)\,, \quad \overline{k^2(t)} = 1 \,,
\end{equation}
where $k(t)$ is a $\tilde{T}$-periodic function,
\begin{equation}\label{k_of_t}
  k(t) = \sum_{n=-\infty}^\infty k_ne^{-i n\tilde{\Omega}t}\,, \quad \tilde{\Omega} = \frac{2\pi}{\tilde{T}} \,,
\end{equation}
for which $k(t)\in\mathbb{R}$ implies $k_{-n}=k_n^*$.
This situation is analogous to an optomechanical system in which the bandwidth of the cavity is much broader that all other characteristic frequencies of the system, except for the optical carrier frequency $\omega_o$ (the bad-cavity approximation). In this case, the Fourier-picture representation of the input-output and Heisenberg-Langevin equations for this system are (see Appendix~\ref{app:eqs_of_motion})
\begin{subequations}\label{eq:badcav_Omega_raw}
  \begin{align}
    \hat{b}^c(\Omega) &= \hat{a}^c(\Omega) \,,\label{eq:cos-IO} \\
    \hat{b}^s(\Omega) &= \hat{a}^s(\Omega) +
      \sqrt{\Gamma}\sum_{n=-\infty}^\infty k_n\hat{X}(\Omega-n\tilde{\Omega}) \,, \label{eq:sin-IO}\\
    \chi^{-1}(\Omega)\hat{X}(\Omega)
      &= \sqrt{\Gamma}\sum_{n=-\infty}^\infty k_n\hat{a}^c(\Omega-n\tilde{\Omega})
        + \hat{f}(\Omega) \,. \label{eq:X-Omega-sol}
  \end{align}
\end{subequations}
Here
\begin{equation}\label{norm_F}
  \hat{f} = \frac{\hat{F}}{\sqrt{\hbar\rho}} \,,
\end{equation}
$\hat{F}$ is the sum of all external forces acting on the oscillator \emph{except} for the QBA, $\hat{a}^{c,s}$ are the amplitude (cosine) and phase (sine) quadratures of the incident light, $\hat{b}^{c,s}$ are the corresponding quadratures of the outgoing light, $\Gamma$ is the mean value (averaged over the period $\tilde{T}$) of the oscillator-light coupling rate,
\begin{equation}\label{chi_norm}
  \chi(\Omega) = \frac{\Omega_0}{\Omega_0^2 -\Omega^2 - 2i\Omega\gamma}
\end{equation}
is the normalized oscillator susceptibility~\footnote{{This form is strictly true for viscous damping $\dot{\hat{P}}=-2\gamma\hat{P}+\cdots$. If instead the oscillator damping acts on both the canonical position and momentum $\dot{\hat{X}}=-\gamma\hat{X}+\cdots$ and $\dot{\hat{P}}=-\gamma\hat{P}+\cdots$ as is typically the case for, e.g., spin oscillators, then a frequency correction term $+\gamma^2$ appears in the denominator of $\chi$. Here we assume the high-$Q$ limit $|\Omega_0|/(2\gamma)\gg 1$ in which this correction (if relevant) is negligible (see also Appendix~\ref{app:chi_eff}).}}, and $\gamma$ is its damping rate (half width at half maximum). The terms $\propto\hat{a}^c(\Omega-n\tilde{\Omega})$ in Eq.~\eqref{eq:X-Omega-sol} represent the net QBA induced by the coupling to the light field, including both fundamental and extraneous contributions. We suppose that the sine quadrature $\hat{b}^s$ is measured by a homodyne detector, in which case the term $\hat{a}^s$ in Eq.~\eqref{eq:sin-IO} represents the imprecision shot noise. 

In the vicinity of the frequencies $\pm\tilde{\Omega}$, using the assumption~\eqref{RWA2}, the mechanical  susceptibility can be approximated as
\begin{equation}\label{chi_nb}
  \chi(\Omega\pm\tilde{\Omega}) \approx \pm\frac{is}{2\ell(\Omega\mp\Lambda)} \,,
\end{equation}
where we have introduced the inverse complex Lorentzian for the oscillator:
\begin{equation}
  \ell(\Omega) = \gamma-i\Omega \,.
\end{equation}

Equations~\eqref{eq:badcav_Omega_raw} are now solved for the signal (sine) quadrature $\hat{b}^s$ of the optical field.
For reference, we provide first the solution for the particular case of an oscillator subject to a constant drive field,  $k(t)=1\Rightarrow k_n=\delta_{n,0}$:
\begin{equation}\label{eq:badcav_Omega_constAmp}
  \hat{{\rm b}}^s(\Omega) = \hat{{\rm a}}^s(\Omega)
  + \sqrt{\Gamma}\chi(\Omega) [\sqrt{\Gamma}\hat{{\rm a}}^c(\Omega) + \hat{f}(\Omega)] \,.
\end{equation}
This equation shows that the QBA is mapped into the output according to the transfer function $\Gamma \chi(\Omega)$. The schemes presented below have essentially the purpose of engineering the effective QBA transfer function with the eventual goal of ensuring cancellation and, thus, a QMFS.

Consider now the case of modulated driving, taking into account the approximations~\eqref{RWA2}. In this case we have for Fourier components $|\Omega|\ll\tilde{\Omega}$ of the output sine quadrature (see Appendix \ref{app:effective}),
\begin{multline}\label{badcav_Omega_var}
  \hat{b}^s(\Omega) = \hat{a}^s(\Omega)
    + \sqrt{\Gamma_{\text{eff}}}\chi_{\text{eff}}(\Omega)
        [\sqrt{\Gamma_{\text{eff}}}\hat{a}^c(\Omega) + \hat{f}_{\text{eff}}(\Omega)]
  \\  + \hat{b}^s_{\rm extra}(\Omega) \,,
\end{multline}
where
\begin{subequations}\label{eq:eff-params}
  \begin{align}
    \Gamma_{\text{eff}} = {} &|k_1|^2\Gamma \,, \label{eq:Gamma-eff}\\
    \chi_{\text{eff}}(\Omega)
      = {}&\frac{\Omega_{\text{eff}}}{\Omega_{\text{eff}}^2+\gamma^2-\Omega^2-2i\gamma\Omega}\,,\label{eq:chi-eff_raw} \\
    \hat{f}_{\rm eff}(\Omega) = {}&\frac{i}{2\Lambda}\bigl[
        e^{-i\Phi}\ell(\Omega+\Lambda)\hat{f}(\Omega+\tilde{\Omega}) \nonumber \\
         &- e^{i\Phi}\ell(\Omega-\Lambda)\hat{f}(\Omega-\tilde{\Omega})
      \bigr] \,,\label{eq:f-eff} \\
    \Phi = {}&\arg k_1 = -\arg k_{-1} \,,\label{eq:Phi}
  \end{align}
\end{subequations}
and
\begin{equation}\label{bs_extra}
  \hat{b}^s_{\rm extra}(\Omega)
  = \frac{is\Gamma_{\text{eff}}}{2|k_1|}\sum_{n\ne0}\biggl[
        \frac{e^{-i\Phi}k_{n+1}}{\ell(\Omega-\Lambda)}
        - \frac{e^{i\Phi}k_{n-1}}{\ell(\Omega+\Lambda)}
      \biggr]\hat{a}^c(\Omega-n\tilde{\Omega})
\end{equation}
is the extraneous back action term originating from the sideband components of the amplitude noise of the incident optical field $\hat{a}^c(\Omega-n\tilde{\Omega})$, converted to the oscillator frequency by means of beating with the modulated optical drive.

Up to the substitutions~(\ref{Omega_eff}, \ref{eq:eff-params}), Eq.~\eqref{badcav_Omega_var} differs from Eq.~\eqref{eq:badcav_Omega_constAmp} in two aspects. The first one is the additional, extraneous back action term~\eqref{bs_extra}. However, contrary to the nominal back action, proportional to $\hat{a}^c(\Omega)$, it commutes with the shot noise $\hat{a}^s(\Omega)$ and can therefore, in principle, be suppressed or completely removed, as we will indeed provide feasible techniques for in Subsec.~\ref{sec:suppression}.
The second aspect is the non-trivial transformation of the external force~\eqref{eq:f-eff}. In particular, let $f$ be a thermal force with the symmetrized spectral density $S_T(\Omega)$, which in the vicinity $\Omega\approx\pm|\Omega_0|$ is approximately constant and equal to
\begin{equation}\label{S_T}
  S_T \approx 2\gamma(2n_T+1) \,,
\end{equation}
where
\begin{equation}
  n_T = \frac{1}{\exp\frac{\hbar|\Omega_0|}{k_BT} - 1}
\end{equation}
is the mean number of thermal quanta, $T$ is the temperature, and $k_B$ is the Boltzmann constant. In this case, the spectral density of $f_{\rm eff}$ is equal to
\begin{equation}\label{S_T_eff_raw}
  S_{T\,{\rm eff}}(\Omega)
  = \frac{ \Omega_{\text{eff}}^2 + \gamma^2 + \Omega^2}{2\Omega_{\text{eff}}^2} S_T \,.
\end{equation}
It is important that this spectral density, while different from Eq.~\eqref{S_T}, has the same order of magnitude and is characterized by the same number of (down-converted) quanta $n_T$.

Equations~(\ref{eq:eff-params}, \ref{bs_extra}) are valid for arbitrary coupling envelopes~\eqref{k_of_t} and thus allows us compare the performance of different envelopes in realizing the effective oscillator. Eq.~\eqref{eq:Gamma-eff} shows that only the $n=\pm1$ Fourier components $k_n$ ($|k_1|=|k_{-1}|$) contribute to the effective readout rate $\Gamma_{\text{eff}}$, whereas the remaining components $k_{n\neq\pm1}$ only contribute to the extraneous QBA~\eqref{bs_extra}. While we present a scheme to completely compensate the extraneous QBA for all stroboscopic-type coupling envelopes ($k_{2n}=0$) in Sec.~\ref{sec:suppression}, it follows from the above that two-tone driving (for which only $k_{\pm1}$ are non-zero) is optimal in the sense that the drive power associated with any non-zero components $k_{n\neq\pm1}$ is wasted for purposes of engineering the effective oscillator.

In the denominator of Eq.~\eqref{eq:chi-eff_raw} and the numerator of Eq.~\eqref{S_T_eff_raw}, the effective frequency $\Omega_{\rm eff}$ appears in the combination $\Omega_{\rm eff}^2+\gamma^2$, playing the role of effective resonance frequency. Typically the required value of the resonance frequency is much larger than $\gamma$, in which case the term $\gamma^2$ can be neglected (the intrinsic bandwidth of state-of-art mechanical resonators can be as small as $\lesssim1\,{\rm mHz}$, whereas in the case of atomic spin oscillators, values down to $\sim3{\rm Hz}$ are feasible~\cite{Balabas_OE_18_5825_2010}).
An important exception, where a minute resonance frequency is required, is the modern GWDs, where the resonance frequency could be as low as $\sim1\,{\rm Hz}$ (as will be treated in Sec.~\ref{sec:free_mass} below). In this case, the value of $\sqrt{\Omega_{\rm eff}^2+\gamma^2}$ in Eq.~\eqref{eq:chi-eff_raw} can be further shifted down by using the virtual rigidity technique considered in Ref.~\cite{19a1ZePoKh}. However, the corresponding term in the effective thermal noise spectral density is not affected by this approach.
In Appendix~\ref{app:chi_eff}, we discuss the physical origin of the term $\gamma^2$ in Eqs.~(\ref{eq:chi-eff_raw}, \ref{S_T_eff_raw}) and show that is can be eliminated using parametric excitation of the down-converted oscillator, giving
\begin{align}
  \chi_{\text{eff}}(\Omega)
    &= \frac{\Omega_{\text{eff}}}{\Omega_{\text{eff}}^2 - \Omega^2 - 2i\gamma\Omega}\,,
    \label{eq:chi-eff} \\
    S_{T\,{\rm eff}}(\Omega)
      &= \frac{\Omega_{\text{eff}}^2 + \Omega^2}{\Omega_{\text{eff}}^2}\gamma(2n_T+1) \,.
      \label{S_T_eff}
\end{align}
We will use these expressions for $\chi_{\text{eff}}$ and $S_{T\,{\text{eff}}}$ henceforth.

\subsection{Suppression of extraneous QBA components}\label{sec:suppression}

Consider now the techniques for suppressing the extraneous QBA~\eqref{bs_extra}. Such QBA, which is not required by the Heisenberg uncertainty relation, is familiar from the stroboscopic and two-tone measurements of a single oscillator quadrature mentioned previously. In the stroboscopic measurement case~\cite{78a1eBrKhVo}, it is the extraneous QBA term that creates the position-measurement precision limit $\Delta x\sim\sqrt{\hbar\tau_{\rm strob}/m}$, where $\tau_{\rm strob}\ll|\Omega_0|^{-1}$ is the measurement duration. In the case of two-tone measurement~\cite{Caves_RMP_52_341_1980}, $\hat{b}^s_{\rm extra}$ is suppressed by a narrowband optical cavity, giving the residual precision limit $\Delta x\sim\sqrt{\hbar\kappa/(m\Omega_0^2)}$, where $\kappa$ is the cavity bandwidth.

Here we combine the benefits of these two schemes generalized to the ``detuned'' setting $\Lambda\neq 0$ in an approach that can be implemented in systems analogous to bad-cavity optomechanics, while avoiding the use of very short optical pulses.
Note that, even though we assume this bad-cavity regime henceforth, the scheme can be straightforwardly extended to work for oscillators embedded in cavities with moderate sideband resolution $\kappa\gtrsim|\Omega_0|$, as this will simply entail a partial (but generally insufficient) suppression of the components in Eq.~\eqref{bs_extra} according to the cavity Lorentzian (insofar as we remain in the weak-coupling regime $\kappa\gg\Gamma$). Finally, in the (technically challenging) good-cavity regime $\kappa\ll|\Omega_0|$ the extraneous QBA is fully suppressed in this manner, rendering additional suppression techniques unnecessary, see, e.g., Refs.~\cite{Caves_RMP_52_341_1980, Woolley_PRA_87_063846_2013}.

\subsubsection{Stroboscopic-type periodic drive\\ and the special case of two-tone drive}

For the purposes of developing our downconversion scheme, we constrain our analysis to the stroboscopic class of periodic drive modulation envelopes,
\begin{equation}
  k(t) = \sum_{n=-\infty}^\infty\left[K\bigl(t-n\tilde{T}-\tau\bigr) - K\bigl(t-(n+\nicefrac12)\tilde{T}-\tau\bigr)\right] \,, \label{eq:k-t_strobo}
\end{equation}
where $K(t)$ is a pulse of duration $\le \tilde{T}/2$ and $\tau$ is an arbitrary time delay.  The envelopes~\eqref{eq:k-t_strobo} are invariant under translation by half a period $t\rightarrow t+\tilde{T}/2$ combined with a sign inversion. Due to the alternating signs of the unit pulses making up the pulse train $k(t)$, all its even Fourier components vanish, $k_{2n}=0$. The restriction~\eqref{eq:k-t_strobo} on $k(t)$ simplifies our analysis somewhat and is warranted in that there is no advantage (from a fundamental viewpoint) in employing modulation envelopes from outside this class.

\begin{figure}[hbt]
\centering
\def\svgwidth{1\columnwidth}
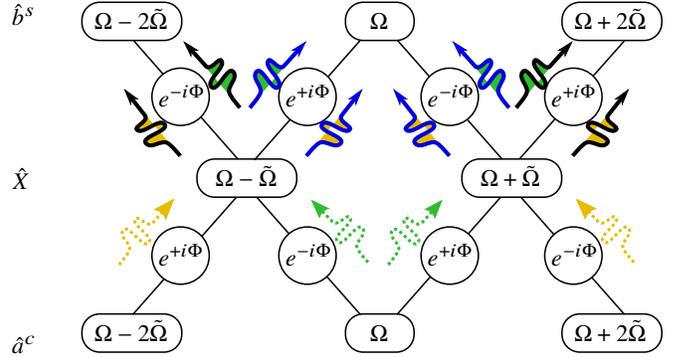
\caption{Spectral mappings for a two-tone-driven oscillator showing the emergence of the extraneous QBA sidebands. Both paths linking the nominal QBA [green dashed arrows] to the nominal output sidebands $\hat{a}^c(\Omega)\rightarrow\hat{b}^s(\Omega)$ [blue arrows with green filling] via $\hat{X}$ have net zero contribution from the phase $\Phi$ in contrast to the extraneous QBA components [yellow dashed arrows] that are mixed into the nominal output $\hat{a}^c(\Omega\pm2\tilde{\Omega})\rightarrow\hat{b}^s(\Omega)$ [blue arrows with yellow filling] with a net phase contribution of $e^{\mp2i\Phi}$. The extraneous output components $\hat{b}^s(\Omega\pm2\tilde{\Omega})$ [black arrows with filling] can be filtered out in post-processing.}
\label{fig:two-tone-scattering-diagram}
\end{figure}

\begin{figure*}[thb]
\centering
\def\svgwidth{0.75\textwidth}
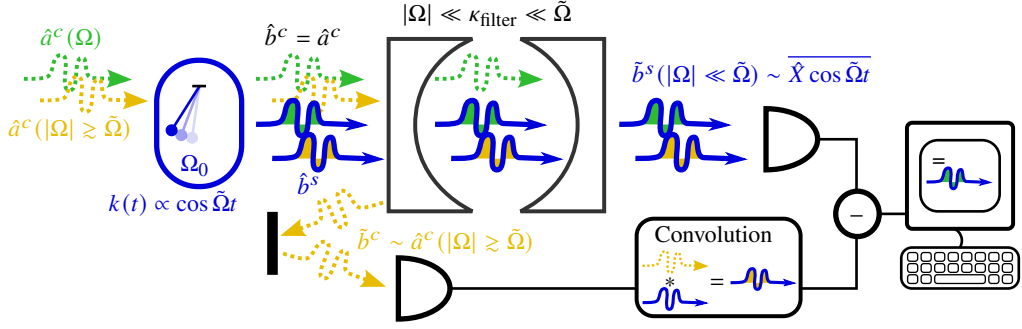
\caption{Scheme for removing extraneous QBA~\eqref{bs_extra} by invoking an auxiliary measurement channel and post-processing. Spectral separation of the output light field $\hat{b}$ from the oscillator is performed using an \emph{external} narrowband cavity ($|\Omega|\ll\kappa_{\text{filter}}\ll\tilde{\Omega}$) resonant at the optical carrier frequency $\omega_{\text{filter}}=\omega_o$. The resulting reflection of the high-frequency sideband components $|\Omega|\gg\kappa_{\text{filter}}$ of $\hat{b}(\Omega)$ (absolute frequencies $\omega_o\pm\Omega$) permits a direct measurement of the stochastic, extraneous QBA force, arising from components $|\Omega|\gtrsim\tilde{\Omega}\gg\kappa_{\text{filter}}$ [yellow, dashed arrow] of the amplitude quadrature $\tilde{b}^c(\Omega)=\hat{a}^c(\Omega)$ [in the particular case of two-tone driving, the relevant components are $\Omega\sim\pm 2\tilde{\Omega}$, see Eq.~\eqref{bs_extra_2}].
The transmitted, low-frequency sideband components $|\Omega|\ll\kappa_{\text{filter}}\ll\tilde{\Omega}$ of $\hat{b}^s(\Omega)$ [blue, thick arrows], Eq.~\eqref{badcav_Omega_var}, are subjected to a standard phase quadrature measurement $\tilde{b}^s(\Omega)=\hat{b}^s(\Omega)$. By post-processing the auxiliary measurement current $\tilde{b}^c$ according to the oscillator response function in Eq.~\eqref{bs_extra} and subtraction from that of $\tilde{b}^s$, the extraneous QBA contribution [blue arrow with yellow filling] is removed from the latter, resulting in an effective oscillator readout of the desired form, Eq.~\eqref{eq:badcav_Omega_constAmp}, containing only the nominal QBA contribution [blue arrow with green filling] induced by $\hat{a}^c(\Omega)$ [green, dashed arrow].}\label{fig:two-tone_plus}
\end{figure*}

Here we present two schemes for suppressing $\hat{b}_{\text{extra}}^s$ that can accommodate arbitrary $k(t)$ of the form~\eqref{eq:k-t_strobo}. However, given the optimality of two-tone probing (in the sense discussed in Subsec.~\ref{subsec:osc-modulated-coup}), we will often use this special case to exemplify our approach in what follows; i.e., we will choose
\begin{equation}
  k_{\pm 1} = e^{\pm i\Phi}/\sqrt{2}
\end{equation}
to be the only non-zero Fourier coefficients of $k(t)$, Eq.~\eqref{k_of_t}, meaning that
\begin{equation}\label{eq:k-t-2tone}
  k(t)=\sqrt{2}\cos(\tilde{\Omega} t - \Phi) \,,
\end{equation}
which is seen to obey the stroboscopic form~\eqref{eq:k-t_strobo} with the unit pulse $K(t)=\sqrt{2}\cos(\tilde{\Omega} t) \Theta(\tilde{T}/4-|t|)$ and the identification $\tau=\Phi/\tilde{\Omega}$ [see Eq.~\eqref{eq:Phi}]. In this case, only two components of $\hat{a}^c$ remain in Eq.~\eqref{bs_extra}:
\begin{equation}\label{bs_extra_2}
  \hat{b}^s_{\rm extra}(\Omega) = \frac{is\Gamma}{4}\biggl[
    \frac{\hat{a}^c(\Omega + 2\tilde{\Omega})}{\ell(\Omega-\Lambda)}e^{-2i\Phi}
    - \frac{\hat{a}^c(\Omega - 2\tilde{\Omega})}{\ell(\Omega+\Lambda)}e^{2i\Phi}
  \biggr] \,.
\end{equation}
The resulting scattering dynamics is sketched in Fig.~\ref{fig:two-tone-scattering-diagram}.

\subsubsection{Scheme \#1 for suppression of unwanted sidebands: Downstream narrowband cavity for separation, \\measurement, and subtraction in post-processing}\label{subsec:measurement-scheme}

Our first scheme for suppression of extraneous QBA is based on a supplementary, direct, and, in principle, perfect measurement of this stochastic force. Since the force is now known, the oscillator's (deterministic) response to it can be calculated and we can achieve a conditional evolution of the effective oscillator in which the extraneous QBA is absent.  This idea has recently been proposed in the context of a QBA-evading \emph{single}-quadrature measurement of a probe system, with the aim of detecting a single phase of a classical force signal~\cite{Vyatchanin_PRA_104_023519_2021}.

For purposes of implementing this idea, suppose that at $|\Omega|\ll\tilde{\Omega}$ the phase quadrature $\hat{b}^s$ is measured, whereas at $|\Omega|\gtrsim\tilde{\Omega}$ the amplitude (QBA) quadrature $\hat{b}^c=\hat{a}^c$ is measured; this can effectively be accomplished using a filter cavity with a bandwidth $\kappa_{\rm filter}$ satisfying the condition (for all $\Omega$ of interest)
\begin{equation}
  |\Omega| \ll \kappa_{\rm filter} \ll \tilde{\Omega} \,,
\end{equation}
see Fig.~\ref{fig:two-tone_plus}. Suppose that the high-frequency part is processed to reproduce the transformation~\eqref{bs_extra}. Subtraction of the result from the measurement record of the low-frequency part, Eq.~\eqref{badcav_Omega_var}, cancels the extraneous QBA term $\hat{b}^s_{\text{extra}}$.
More generally, for a finite efficiency $\eta_{\rm{aux}}$ of the auxiliary detector measuring $\hat{b}^c$, the noise spectral density of $\hat{b}^s_{\rm extra}$~\eqref{bs_extra} can be suppressed by the factor $1-\eta_{\text{aux}}$, as follows from a straightforward optimization of the gain with which the auxiliary measurement is combined with the primary measurement.
Note that the transformation~\eqref{bs_extra} is causal, and therefore this operation can be performed almost in real time (up to the short delay $\sim1/\kappa_{\rm filter}$ imposed by the filter cavity).

\subsubsection{Scheme \#2 for suppression of unwanted sidebands:\\ Twin oscillators two-tone driven out of phase}\label{subsec:coh-scheme}

\begin{figure*}[t]
\centering
\def\svgwidth{0.75\textwidth}
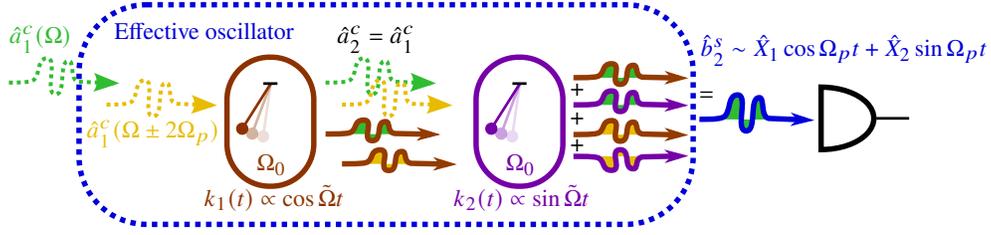
\caption{Scheme for removing the extraneous QBA contribution~\eqref{bs_extra_2} [thick arrows with yellow filling] induced by the spectral sideband components $\sim\pm 2\tilde{\Omega}$ [yellow dashed arrow] of $\hat{a}^c_1$ using coherent cancellation. A single, effective oscillator with the desired behavior is formed from cascading two identical oscillators that are two-tone-driven out of phase: The extraneous QBA responses [thick arrows with yellow filling] cancel, while the nominal QBA responses [thick arrows with green filling] interfere constructively.}
\label{fig:scheme2}
\end{figure*}

As an alternative to the measurement-based scheme laid out in the previous subsection, we here propose a scheme based on coherent cancellation of the extraneous QBA. The simplest case for the present scheme is that of two-tone probing introduced above and, given the optimality of this probing type, we will mostly focus on this here (its generalization to other forms of $K(t)$ is discussed at the end of this subsection).
In this case, the cancellation of the extraneous QBA is achieved by invoking a second oscillator subject to periodic driving with the same unit pulse $K(t)$ but a different time delay $\tau=\Phi/\tilde{\Omega}$~[see Fig.~\ref{fig:scheme2} and Eqs.~\eqref{eq:k-t_strobo}--\eqref{eq:k-t-2tone}].

To this end, consider two oscillators $j\in\{1,2\}$ with the same parameter values $\Gamma_{\text{eff},j}=\Gamma_{\text{eff}}/2$, $\gamma_j=\gamma$,
and $\Omega_{0,j}=\Omega_0$ and which are periodically driven according to $k_j(t)=\sqrt{2}\cos(\tilde{\Omega} t - \Phi_j)$ with $\Phi_1=0$ and $\Phi_2=\pi/2$, respectively. It follows from Eq.~\eqref{bs_extra_2} that in this case,
\begin{equation}
  \hat{b}^s_{\rm extra\,1} = -\hat{b}^s_{\rm extra\,2} \,,
\end{equation}
whence these terms can be made to interfere destructively by combining the two oscillators in cascade. Indeed, identifying the output from system 1 with the input to system 2, $\hat{a}_{2}^{c,s}(\Omega)=\hat{b}_{1}^{c,s}(\Omega)$, the resulting output fields from the second system are seen to be $\hat{b}^c_2=\hat{a}^c_1$ and (valid for $|\Omega|\ll \tilde{\Omega}$)
\begin{equation}\label{eq:b-s_coh-comb}
  \hat{b}_2^s(\Omega)
           = \hat{a}_1^s(\Omega)  + \sqrt{\Gamma_{\text{eff}}}\chi_{\text{eff}}(\Omega)[\sqrt{\Gamma_{\text{eff}}} \hat{a}^c_1(\Omega)
      + \hat{f}_{\text{eff}}^{\prime}(\Omega)] \,,
\end{equation}
where the joint force operator is $\hat{f}_{\text{eff}}^{\prime}(\Omega)\equiv\{\hat{f}_{\text{eff},1}(\Omega)+\hat{f}_{\text{eff},2}(\Omega)\}/\sqrt{2}$.
If the $\hat{f}_{\text{eff},j}$ are in a thermal state, the $\hat{f}_{\text{eff}}^\prime$ will obviously have the same spectral properties as the original $\hat{f}_{\text{eff},j}(\Omega)$ if the thermal bath temperatures are equal (in addition to the previous assumption of identical oscillator parameters). Note that in Eq.~\eqref{eq:b-s_coh-comb} the nominal QBA responses $\propto\hat{a}^c(\Omega)$ of the two oscillators have interfered constructively.
Comparison between Eq.~(\ref{eq:b-s_coh-comb}) and the constant-drive analog, Eq.~\eqref{eq:badcav_Omega_constAmp}, shows that we have recovered
the desired transformation, i.e., without admixture of the extraneous QBA at $\Omega\pm2\Omega_{p}$ contained in $\hat{b}_{\text{extra}}^{s}$. The net, mean drive power $\propto\sum_j\Gamma_{\text{eff},j}=\Gamma_{\text{eff}}$ equals that of the measurement-based scheme presented in the previous paragraph.

To clarify the nature of the joint, effective oscillator that emerges from the combination of individual oscillators 1 and 2, consider its time-domain input-output relation for the present choice of $k_j(t)$ and $\Phi_j$~[cf.\ Eq.~\eqref{eq:sin-IO}]
\begin{align}\label{eq:coh-cancel-IO}
  \hat{b}^s(t)
  ={}&\hat{a}^s(t) + \sqrt{\frac{\Gamma_{\text{eff}}}{2}}
    [\hat{X}_1(t)\cos\tilde{\Omega}t + \hat{X}_2(t)\sin\tilde{\Omega}t] \nonumber& \\
  \approx{}& \hat{a}^s(t) + \sqrt{\Gamma_{\text{eff}}}  \Bigg[
      \frac{\hat{\mathcal{X}}_1(t)+\hat{\mathcal{P}}_2(t)}{\sqrt{2}}\cos \Omega_{\text{eff}} t &\nonumber\\
      &\hspace{2cm}+ \frac{\hat{\mathcal{P}}_1(t)-\hat{\mathcal{X}}_2(t)}{\sqrt{2}}\sin \Omega_{\text{eff}} t \Bigg] ,
\end{align}
where in the last line we have, in complete analogy to Eq.~\eqref{eq:two-tone-rot}, averaged out the fast oscillations at $\sim2\tilde{\Omega}$ and introduced the slowly varying quadratures $(\hat{\mathcal{X}}_j,\hat{\mathcal{P}}_j)$, Eq.~\eqref{eq:slowly-rot-vars}, defined with respect to the frequency $\Omega_0$.
Equation~\eqref{eq:coh-cancel-IO} makes it clear that we have realized a single effective oscillator with effective frequency $\Omega_{\text{eff}}=s\Lambda$ and canonical, slowly-varying quadratures $[(\hat{\mathcal{X}}_1+\hat{\mathcal{P}}_2)/\sqrt{2},\, (\hat{\mathcal{P}}_1-\hat{\mathcal{X}}_2)/\sqrt{2}]=i$ [contrasting the EPR oscillator described by Eqs.~\eqref{eq:EPR-osc} and~\eqref{eq:EPR-comm}].
This joint oscillator can then, in turn, be combined with other systems to form a QMFS.

The coherent cancellation scheme presented here can, in principle, be extended to arbitrary stroboscopic-type envelopes $k(t)$ (i.e., which have other non-zero Fourier coefficients than $k_{\pm1}$). For such pulses, even Fourier components vanish $k_{2j}=0$ (as noted previously), and hence the extraneous QBA response~\eqref{bs_extra} can only have contributions for even $n$ ($\neq0$). It is straightforward to show that in the generalized scenario of a cascade with $N$ identical oscillators driven stroboscopically by the same unit pulse $K(t)$, but with different time delays $\tau_l=\pi(l-1)/(N\tilde{\Omega})$ where $1\leq l \leq N$, all extraneous QBA components in Eq.~\eqref{bs_extra} obeying $n/2\, \text{mod}\, N \neq 0$ are cancelled, whereas terms $n/2\, \text{mod}\, N = 0$ interfere constructively. Hence, it follows from Eq.~\eqref{bs_extra} that $N$ must equal half the number of non-zero Fourier components $k_n$ plus one in order for the extraneous QBA to be cancelled entirely.

\section{Applications}\label{sec:applications}

A number of important applications of QBA evasion, and hence of QMFSs, have already been established in the literature. These include pulsed and continuous force sensing below the SQL~\cite{Wasilewski_PRL_104_133601_2010,Tsang_PRL_105_123601_2010,Tsang_PRX_2_031016_2012,Woolley_PRA_87_063846_2013,Zhang_PRA_88_043632_2013, Bariani_PRA_92_043817_2015, Motazedifard_NJP_18_073040_2016, 18a1KhPo, Mason_NPhys_15_745_2019}, and entanglement generation between (potentially) distant nodes in a quantum network~\cite{Hammerer_PRL_102_020501_2009, Thomas_NPhys_17_228_2021, Lepinay_Science_372_625_2021}. Our approach provides the means to implement these important protocols in a hybrid system of otherwise spectrally incompatible subsystems.

Here we will review some of these protocols, beginning with entanglement generation and impulse sensing in Subsec.~\ref{subsec:ent-gen}. Next, we discuss continuous force sensing in Subsec.~\ref{subsec:stat-force-sens}. In this connection, we provide estimates for the application of our method to circumventing the quantum noise bottleneck in state-of-the-art GW interferometers in Subsec.~\ref{sec:estimates}.

In addition to the effective resonance frequency $\Omega_{\text{eff}}$, other (effective) parameters characterizing the oscillator are its coupling rate to the itinerant field $\Gamma_{\text{eff}}$, its decay rate $\gamma$, and the associated thermal noise temperature. As we will see below, the roles and importance of these parameters depend on the application at hand.

\subsection{Entanglement generation and impulse sensing}\label{subsec:ent-gen}

The simultaneous measurement of a pair of commuting EPR variables $(\hat{\mathcal{X}}_{\text{EPR}},\hat{\mathcal{P}}_{\text{EPR}})\equiv(\hat{\mathcal{X}}_1+\hat{\mathcal{X}}_2,\,\hat{\mathcal{P}}_1-\hat{\mathcal{P}}_2)/\sqrt{2}$, as discussed in connection with Eq.~\eqref{eq:EPR-osc}, will (conditionally) project them into a completely well-defined state $\text{Var}[\hat{\mathcal{X}}_{\text{EPR}}],\text{Var}[\hat{\mathcal{P}}_{\text{EPR}}]\rightarrow 0$ in the limit where the measurement rate overwhelms thermal decoherence and measurement imprecision shot noise.
For Gaussian systems and measurements, on which we focus throughout this paper, the best estimates of $(\hat{\mathcal{X}}_{\text{EPR}},\hat{\mathcal{P}}_{\text{EPR}})$ can be extracted from the measurement current if the input-output relations~\eqref{eq:sin-IO} and system dynamics~\eqref{eq:X-Omega-sol} are known, using, e.g., Wiener filtering (see for instance Ref.~\cite{Thomas_NPhys_17_228_2021}). If the simultaneous estimation of the EPR oscillator variables $[\hat{\mathcal{X}}_{\text{EPR}},\hat{\mathcal{P}}_{\text{EPR}}]=0$ attains an accuracy beyond what is possible for canonical oscillator variables $[\hat{X},\hat{P}]=i$,
this implies entanglement between the subsystems constituting the EPR oscillator according to the Duan inseparability criterion~\cite{Duan_PRL_84_2722_2000}
\begin{equation}\label{eq:Duan}
\Sigma_{\text{EPR}}\equiv\text{Var}[\hat{\mathcal{X}}_{\text{EPR}}]+\text{Var}[\hat{\mathcal{P}}_{\text{EPR}}] < 1\,,
\end{equation}
which is necessary and sufficient in the present case of Gaussian states.
Seeing as the purpose of our scheme is exactly to engineer the required QBA-evading measurements of such EPR pairs formed from disparate subsystems (see Fig.~\ref{fig:topologies}), it is a well suited means for entanglement generation in hybrid quantum networks. This is an essential resource, e.g., for the teleportation of quantum states.

The ability to prepare a well-defined EPR state also permits the sensing of an impulse signal without limitations imposed by quantum noise.
After the preparation of the well-defined state ($\Sigma_{\text{EPR}}\approx 0$), the measurement is turned off and a pulsed signal force acts on the EPR oscillator, causing the net displacement $(\Delta\hat{\mathcal{X}}_{\text{EPR}},\Delta\hat{\mathcal{P}}_{\text{EPR}})$. Performing now a second efficient QBA-evading measurements allows us to retrodict the displacement with a precision only limited by the thermal decoherence incurred while the signal force was acting. For an imperfect joint measurement ($\Sigma_{\text{EPR}}>0$), the residual quantum noise variance contribution to the sensing of either of $(\Delta\hat{\mathcal{X}}_{\text{EPR}},\Delta\hat{\mathcal{P}}_{\text{EPR}})$ from imperfect preparation and retrodiction is $\approx\Sigma_{\text{EPR}}$ in the regime where the rotating-wave approximation (RWA) in the light-oscillator coupling is valid, $|\Omega_i|\gg\Gamma_i,\gamma$ (leading to $\text{Var}[\hat{\mathcal{X}}_{\text{EPR}}]\approx \text{Var}[\hat{\mathcal{P}}_{\text{EPR}}]$), and the classical cooperativity is large $\Gamma_i/\gamma_i\gg 1$ (rendering the amplitude decay negligible over the effective measurement duration); see, e.g., the SM of Ref.~\cite{Rossi_PRL_123_163601_2019}.

We will now review how $\Sigma_{\text{EPR}}$, Eq.~\eqref{eq:Duan}, depends on the (effective) oscillator parameters in order to establish criteria for efficient entanglement and sensing performance. Methods for determining the conditional value of $\Sigma_{\text{EPR}}$ for Gaussian systems and measurements can be found in, e.g., Ref.~\cite{Woolley_PRA_87_063846_2013, Thomas_NPhys_17_228_2021, Cernotik_PRA_92_012124_2015} and will not be discussed here.
For specificity, we consider the probing of two (``bare'' or effective) counter-rotating oscillators [Eq.~\eqref{eq:counter-rot-cond}] jointly measured using the linear topology. We allow for a finite (power) transmission between the oscillators $\nu$ and a finite detection efficiency $\eta$ of the light field after interaction with both subsystems (these losses amount to placing beam splitters with transmission parameters $\nu$ and $\eta$, respectively, between and after the subsystems in Fig.~\ref{fig:topologies}(top)). We focus again on the RWA and high-classical-cooperativity regime; this implies, in particular, that we can take $S_{T,{\text{eff}}}(\Omega)\approx S_T$, Eq.~\eqref{S_T_eff}. In this case, the EPR variances entering $\Sigma_{\text{EPR}}$ will be QBA-free if the readout rates are matched $\Gamma_1=\Gamma_2$ and we have negligible optical losses between the subsystems $\nu\approx1$; assuming these conditions are fulfilled it can be found that~\cite{Woolley_PRA_87_063846_2013}
\begin{equation}\label{eq:V-c_thermal}
\Sigma_{\text{EPR}} \approx \frac{1}{2\sqrt{\eta}} \sqrt{\frac{1}{C_{q,1}}+\frac{1}{C_{q,2}}}\,,
\end{equation}
where we have introduced the quantum cooperativity $C_{q,i}\equiv (\Gamma_i/2)/S_{T,i}$, i.e., the ratio of (nominal) QBA to (effective) thermal noise spectral densities as they enter Eq.~\eqref{eq:badcav_Omega_constAmp} or~\eqref{badcav_Omega_var}, as the case may be. We hence see from Eq.~\eqref{eq:V-c_thermal} that in order to have entanglement of the two oscillators according to the criterion~\eqref{eq:Duan}, we must have $C_{q,i}\gtrsim 1/(2\eta)$. Hence the relevant experimental regime is that of quantum cooperativities of at least order unity, which has already been achieved in several relevant platforms~\cite{Thomas_NPhys_17_228_2021,Lepinay_Science_372_625_2021}. The inevitable presence of optical losses between distant subsystems $\nu<1$ imposes the lower bound~\cite{Huang_PRL_121_103602_2018}
\begin{equation}\label{eq:V-c_opt}
\Sigma_{\text{EPR}} > \frac{1}{\sqrt{\eta}}\sqrt{\frac{1-\nu}{1+3\nu}}\,.
\end{equation}
Notably, this lower bound is less than unity [the entanglement limit~\eqref{eq:Duan}] for any amount of intersystem losses $0\leq\nu\leq1$ provided that detection is perfect $\eta=1$. Even for a modest transmission of $\nu\sim0.45$, the optical loss bound~\eqref{eq:V-c_opt} allows $\Sigma_{\text{EPR}}\gtrsim1/2$, i.e., EPR entanglement of $\sim3\text{dB}$ (assuming $\eta\sim1$).
Equations~\eqref{eq:V-c_thermal} and~\eqref{eq:V-c_opt} capture the bottlenecks imposed by thermal noise and optical losses, respectively, for the QMFS applications of conditional entanglement generation and impulse sensing described here.
Entanglement generation in the two-oscillator cascade considered here can also be achieved unconditionally (without measurement) by invoking non-local dynamical back-action effects for the EPR oscillator $(\hat{\mathcal{X}}_{\text{EPR}},\hat{\mathcal{P}}_{\text{EPR}})$~\cite{Huang_PRL_121_103602_2018}.

\subsection{QBA evasion in stationary force sensing}\label{subsec:stat-force-sens}

\subsubsection{Assumptions and approximations}

We here consider how to engineer QBA-evading continuous detection of a force acting on a probe system. Hybrid optomechanical schemes based on the serial and parallel topologies (see Fig.~\ref{fig:topologies}) were considered, respectively, in Refs.~\cite{Moeller_Nature_547_191_2017} and~\cite{18a1KhPo} for the case where the negative-mass reference frame is implemented by an atomic spin ensemble with negative evolution frequency. Here we extend those treatments to a much wider class of oscillators.

To aide the exposition of the present application, we denote the parameters of the first and second subsystems (previously `1' and `2') by the subscripts $P$ (probe) and $A$ (auxiliary), respectively.
We assume that the signal force acts on the probe oscillator and allow the auxiliary oscillator to be an effective, down-converted one with the effective parameters defined by Eqs.~\eqref{eq:eff-params} and with the extra noise~\eqref{bs_extra} compensated as discussed in Sec.~\ref{sec:suppression}. 
For simplicity, we do not take into account here the optical losses, which were considered in our previous works~\cite{18a1KhPo, 19a1ZePoKh} and which are not affected by the down-conversion.

The corresponding input-output relations for the two subsystems [see Eqs.~(\ref{badcav_Omega_var}, \ref{eq:eff-params})] are the following:
\begin{subequations}\label{io_PA}
  \begin{align}
    \hat{b}^s_P(\Omega) ={}& \hat{a}^s_P(\Omega) \\
      &+ \sqrt{\Gamma_P}\chi_P(\Omega)[
          \sqrt{\Gamma_P}\hat{a}^c_P(\Omega) + \hat{f}_P(\Omega) + f_{\rm sig}(\Omega)] \,, \nonumber\\
    \hat{b}^s_A(\Omega) ={}& \hat{a}^s_A(\Omega) \\
      &+ \sqrt{\Gamma_{A\,{\rm eff}}}\chi_{A\,{\rm eff}}(\Omega)[
            \sqrt{\Gamma_{A\,{\rm eff}}}\,\hat{a}_A^c(\Omega) + \hat{f}_{A\,{\rm eff}}(\Omega)
          ] \,,\nonumber
  \end{align}
\end{subequations}
where $f_{\rm sig}$ is the normalized signal force, see Eq.~\eqref{norm_F}, $\hat{f}_P$ is the thermal noise of the probe oscillator, and $\hat{f}_{A\,\rm eff}$ is the effective thermal noise of the auxiliary system with the spectral density $S_{T\,{\rm eff}}(\Omega)$, see Eq.~\eqref{S_T_eff}.

\subsubsection{Serial topology}

Consider first the more simple serial topology. To be specific, we suppose, that in the scheme of Fig.~\ref{fig:topologies}(top), the probe subsystem goes first, followed by the down-converted auxiliary one (so-called {\it post-filtering}); but actually, the result does not depend on the ordering. In this case, Eqs.~\eqref{io_PA} have to be supplemented by the following ones:
\begin{equation}
  \hat{a}^{c,s}_A = \hat{b}^{c,s}_P \,,
\end{equation}
giving that
\begin{equation}\label{out_s}
  \hat{b}^s_A
  = \sqrt{\Gamma_P}\chi_P(\Omega)
      [f_{\rm sig}(\Omega) + \hat{f}_{\rm sum}(\Omega) + \hat{f}_P(\Omega)] \,,
\end{equation}
where
\begin{multline}\label{f_sum_s}
  \hat{f}_{\rm sum}(\Omega) = \frac{\chi_P^{-1}(\Omega)}{\sqrt{\Gamma_P}}\hat{a}^s_P(\Omega)\\
    + \sqrt{\Gamma_{A\,\rm eff}}\chi_{A\,\rm eff}(\Omega)\biggl[
          K_{\rm res}(\Omega)\hat{a}^c_P
          + \frac{\chi_P^{-1}(\Omega)}{\sqrt{\Gamma_P}}\hat{f}_{A\,{\rm eff}}(\Omega)
        \biggr]
\end{multline}
is the total quantum noise and
\begin{equation}\label{K_res}
  K_{\rm res}(\Omega)
    = \sqrt{\frac{\Gamma_{A\,\rm eff}}{\Gamma_P}}\chi_P^{-1}(\Omega)
      + \sqrt{\frac{\Gamma_P}{\Gamma_{A\,\rm eff}}}\chi_{A\,\rm eff}^{-1}(\Omega) \,.
\end{equation}
Note that the probe thermal noise $\hat{f}_{P}$ appears in Eq.~\eqref{out_s} [as well as in Eq.~\eqref{f_sum_p} below] as an additional, uncorrelated term and therefore can be factored out from the present consideration.

We assume that the input light is prepared in a squeezed state with the logarithmic squeeze factor $r$ and with the squeeze angle equal to zero, which corresponds to the spectral densities of the quadratures $\hat{a}_P^{c,s}$ equal to
\begin{equation}
  S_c = \frac{e^{2r}}{2} \,,\quad S_s = \frac{e^{-2r}}{2} \,.
\end{equation}
In this case, the spectral density of the noise \eqref{f_sum_s} is equal to
\begin{multline}\label{S_f_sum_s}
  S^f_{\rm ser} =  \frac{1}{2}\biggl\{
      \frac{|\chi_P^{-1}(\Omega)|^2}{\Gamma_P}e^{-2r}
      \\+ \Gamma_{A\,\rm eff}|\chi_{A\,\rm eff}(\Omega)|^2\biggl[
            |K_{\rm res}(\Omega)|^2e^{2r}
            + 2\frac{|\chi_P^{-1}(\Omega)|^2}{\Gamma_P|\Omega_{A\,\rm eff}|}
              \tilde{S}_T(\Omega)
          \biggr]
    \biggr\} .
\end{multline}
Here the term $|K_{\rm res}(\Omega)|^2e^{2r}$ corresponds to the residual back action noise and
\begin{equation}\label{tilde_S_T}
  \tilde{S}_T(\Omega) \equiv |\Omega_{A\,{\rm eff}}|S_{T}(\Omega)
    = \frac{\Omega_{A\,\rm eff}^2 + \Omega^2}{|\Omega_{A\,\rm eff}|}\gamma(2n_T+1) \,.
\end{equation}

\subsubsection{Parallel topology}

In the case of the parallel topology [Fig.~\ref{fig:topologies}(bottom)], the combined signal-normalized output current from the probe and auxiliary, Eqs.~\eqref{io_PA}, can be presented as follows:
\begin{align}
  \tilde{f}_{\rm sum}(\Omega)
  &= \frac{\hat{b}^s_P(\Omega)}{\sqrt{\Gamma_P}\chi_P(\Omega)}
    + \frac{\alpha(\Omega)\hat{b}^s_A(\Omega)}
        {\sqrt{\Gamma_{A\,\rm eff}}\chi_{A\,\rm eff}(\Omega)}\nonumber \\
  &= f_{\rm sig}(\Omega) + \hat{f}_{\rm sum}(\Omega) + \hat{f}_{P}(\Omega) \,,\label{f_sum_p}
\end{align}
where
\begin{equation}
  \hat{f}_{\rm sum}(\Omega)
  = \hat{f}_{P\,\rm sum}(\Omega) + \alpha(\Omega)\hat{f}_{A\,\rm sum}(\Omega)
\end{equation}
is the total quantum noise, whereas $\hat{f}_{P\rm sum}$ and $\hat{f}_{\rm A\,\rm sum}$ are the sum noises of the respective channels:
\begin{subequations}
  \begin{align}
    \hat{f}_{P\,\rm sum}(\Omega)
      &= \frac{\chi_P^{-1}(\Omega)}{\sqrt{\Gamma_P}}\hat{a}^s_P(\Omega)
        + \sqrt{\Gamma_P}\hat{a}_P^c(\Omega) \,, \\
    \hat{f}_{A\,\rm sum}(\Omega)
      &= \frac{\chi_{A\,\rm eff}^{-1}(\Omega)}{\sqrt{\Gamma_{A\,\rm eff}}}
          \hat{a}^s_A(\Omega)
        + \sqrt{\Gamma_{A\,\rm eff}}\,\hat{a}_A^c(\Omega)
        + \hat{f}_{A\,{\rm eff}}(\Omega) \label{eq:f-A}\,,
  \end{align}
\end{subequations}
and $\alpha(\Omega)$ is the frequency-dependent relative weight factor in post-processing (to be optimized).

We assume that the squeeze angle of the input two-mode squeezed light is equal to zero, which corresponds to the spectral densities of all four input quadratures equal to
\begin{subequations}\label{eq:entanglement-corr}
  \begin{gather}
    S_{{\rm a}_{P,A}^{c,s}} = \frac{\cosh2r}{2} \,,\label{eq:entanglement-corr_a} \\
    \intertext{and the only non-vanishing components of the cross-correlation matrix equal to}
    S_{{\rm a}_P^c{\rm a}_A^c} = -S_{{\rm a}_P^s{\rm a}_A^s} = \frac{\sinh2r}{2} \,,
      \label{cross_corr}
  \end{gather}
\end{subequations}
where $r$ is the squeeze factor. In this case, the spectral density of $\hat{f}_{\rm sum}$, optimized with respect to $\alpha$, is equal to (see Sec.~IIIB of Ref.~\cite{19a1ZePoKh})
\begin{gather}\label{S_f_sum_p}
  S_{\rm par}^f(\Omega) = \frac{1}{2\Omega_P} \big(
      K_P(\Omega)[K_{A\,\rm eff}(\Omega) + 2\tilde{S}_{T}(\Omega)\cosh2r] \\
      + \Omega_P|\Omega_{A\,\rm eff}||K_{\rm res}(\Omega)|^2\sinh^2 2r \big)\big/
  \big(K_{A\,\rm eff}(\Omega)\cosh2r + 2\tilde{S}_{T}(\Omega)\big) \,,\nonumber
\end{gather}
where
\begin{subequations}\label{eq:K-defs}
  \begin{align}
    K_P(\Omega) &= \frac{|D_P(\Omega)|^2}{\Gamma_P\Omega_P} + \Gamma_P\Omega_P \,, \\
    K_{A\,\rm eff}(\Omega)
      &= \frac{|D_{A\,{\rm eff}}(\Omega)|^2}{\Gamma_{A\,{\rm eff}}|\Omega_{A\,{\rm eff}}|}
        + \Gamma_{A\,{\rm eff}}|\Omega_{A\,{\rm eff}}| \,,
  \end{align}
\end{subequations}
and
\begin{subequations}
  \begin{gather}
    D_P(\Omega) = \Omega_P^2 - \Omega^2 - 2i\gamma_P\Omega \,, \\
    D_{A\,\rm eff}(\Omega) = \Omega_{A\,\rm eff}^2 - \Omega^2 - 2i\gamma_A\Omega \,,
  \end{gather}
\end{subequations}
are the rescaled response functions of the probe and effective auxiliary subsystems.

Here, as in the serial case, the residual back action noise term [the last one in the numerator of Eq.~\eqref{S_f_sum_p}] is proportional to $|K_{\rm res}|^2$ and scales with $r$ as $\sinh^2 2r/\cosh 2r\propto e^{2r}$ for large $r$.

\subsubsection{Cancellation of QBA}

It can be seen from the two previous subsections, that in order to eliminate the QBA, the following condition has to be fulfilled for all signal frequencies of interest:
\begin{equation}\label{QMFS1}
  K_{\rm res}(\Omega) = 0 \,.
\end{equation}
It is easy to see that it is equivalent to matching the (effective) susceptibilities of the two subsystems as
\begin{equation}\label{QMFS2}
  \Gamma_{A\,\rm eff}\chi_{A\,\rm eff}(\Omega) + \Gamma_P\chi_P(\Omega) = 0 \,,
\end{equation}
in accordance with the QMFS approach of ensuring destructive interference between the QBA transfer functions of the subsystems [see remark below Eq.~\eqref{eq:badcav_Omega_constAmp}]. On account of Eqs.~(\ref{chi_norm}, \ref{eq:chi-eff}), the requirement~\eqref{QMFS2} translates to the following three Fourier-frequency-independent conditions:
 \begin{subequations}\label{K_res_zero-cond}
  \begin{gather}
     \Gamma_{A\,\rm eff}\Omega_{A\,\rm eff} + \Gamma_P\Omega_P = 0 \,,\label{eq:QBA-match}\\
     \Omega_{A\,\rm eff}^2 = \Omega_P^2\,, \label{Omega-match}\\
  \gamma_{A\,\rm eff} = \gamma_P \,.\label{Im_K_res}
  \end{gather}
\end{subequations}

Substituting these matching conditions into Eqs.~(\ref{S_f_sum_s}, \ref{S_f_sum_p}) and normalizing the resulting spectral densities to the physical (dimensional) force units, 
\begin{equation}\label{eq:S^F-conv}
S_{\rm ser,\,par}^F = \hbar\rho_P S_{\rm ser,\,par}^f\,, 
\end{equation}
we obtain the following spectral densities with completely suppressed QBA,
\begin{align}
  S_{\rm ser}^F(\Omega) 
    &= \frac{\hbar m}{2}\biggl[
        \frac{|D_P(\Omega)|^2}{\Gamma_P\Omega_P}e^{-2r} + 2\tilde{S}_T(\Omega)
      \biggr] , \label{eq:S-F-ser_opt}\\
  S_{\rm par}^F(\Omega) 
    &= \frac{\hbar m}{2}\frac{
          K_P(\Omega)[K_P(\Omega) + 2\tilde{S}_{T}(\Omega)\cosh2r]
        }{K_P(\Omega)\cosh2r + 2\tilde{S}_{T}(\Omega)} \,. \label{eq:S-F-par_opt}
\end{align}
It has to be noted, that while the down-conversion approach can engineer the fulfillment of the first two conditions (canceling the real part of $K_{\rm res}$), the matching of the damping rates~\eqref{Im_K_res} can be problematic in general. However, the corresponding residual term scales as $1/Q_P^2,1/Q^2_{A\,\rm{eff}}\ll1$ in terms of the $Q$-factors of the probe and effective auxiliary systems:
\begin{equation}
  (\Im K_{\rm res})^2 = \Omega^2\biggl(\frac{1}{Q_P} - \frac{1}{Q_{A\,\rm{eff}}}\biggr)^2 .
\end{equation}
It should also be noted that while the conditions~\eqref{K_res_zero-cond} ensure perfect QBA cancellation, they will not in general lead to the exact optimum of the sensitivity, due to the presence of the auxiliary thermal noise~\cite{20a1KhZe}.

\subsubsection{Engineering a free negative mass}\label{sec:free_mass}

The special case of a very-low-frequency probe oscillator deserves special consideration for two reasons. First, if $\Omega_P\to0$, then the condition~\eqref{Omega-match} cannot be strictly satisfied (without invoking, e.g., the additional down-conversion mechanism of virtual rigidity~\cite{19a1ZePoKh}). Second, this case is relevant to broadband off-resonant force sensing in the frequency band well above the resonance frequency $\Omega_P$. A well known and very important example is the laser GWDs, which use very low-frequency ($\Omega_P/2\pi\sim1\,{\rm Hz}$) pendulums as probe objects. This frequency is much smaller than all other characteristic frequencies of the GWDs, including the lower bound of their sensitivity band $\Omega_{\rm low}$. Therefore, to good approximation, these pendulums behave like free (positive) masses with susceptibility~\footnote{The frequency $\Omega_P$  still appear in Eq.~\eqref{chi_P_fm} due to our use of the normalized position $X$ and force $f$, where $\Omega_P$ is absorbed in $\rho_P$. In the final (unnormalized) equations, they appear only within the product $\Gamma_P\Omega_P$, which is independent of $\Omega_P$.}
\begin{equation}\label{chi_P_fm}
  \chi_P(\Omega) = -\frac{\Omega_P}{\Omega^2}\,.
\end{equation}

In principle, there are no fundamental reasons precluding the auxiliary effective frequency from being reduced to arbitrarily small values. However, this could cause two problems. First, in the frequency band $\Omega>|\Omega_{A\,\rm eff}|$, the effective thermal noise spectral density~\eqref{tilde_S_T} increases with the decrease of $|\Omega_{A\,\rm eff}|$. Second, for a given value of $\Gamma_{A\,\rm eff}$ (which does not depend on $|\Omega_{A\,\rm eff}|$) the effective coupling factor $\Gamma_{A\,\rm eff}|\Omega_{A\,\rm eff}|$ goes to zero if $|\Omega_{A\,\rm eff}|\to0$. In principle, the smallness of $|\Omega_{A\,{\rm eff}}|$ required by Eq.~\eqref{Omega-match} must be compensated by a sufficiently large value of $\Gamma_{A\,{\rm eff}}$~[Eq.~\eqref{eq:Gamma-eff}] in order to fulfill the QBA strength matching condition~\eqref{eq:QBA-match}, which could be demanding in practice.

Due to these reasons, in the case of $\Omega_P\to0$ it is useful to formulate a pragmatic criterion for observing QBA reduction in experiment even if it falls short of fulfilling the ideal condition in Eq.~\eqref{Omega-match}:
\begin{equation}\label{eq:Omega-low-cond}
  |\Omega_{A\,\rm eff}| < \Omega_{\rm low} \,.
\end{equation}
Alternatively, the requirement~\eqref{Omega-match} can be fulfilled by choosing a larger initial value of $|\Omega_{A\,{\rm eff}}|$ and combining the down-conversion scheme presented here with an additional frequency shift by means of the virtual rigidity effect~\cite{19a1ZePoKh}.

In order to account for imperfect matching $\Omega_{A\,\rm eff}^2\neq \Omega_P^2$ and $\gamma_{A\,\rm eff}\neq \gamma_P$ in our estimates below, we substitute only the matching condition~\eqref{eq:QBA-match} into Eqs.~(\ref{S_f_sum_s}, \ref{S_f_sum_p}); we obtain, in the normalization~\eqref{eq:S^F-conv},
\begin{align}
  S_{\rm ser}^F(\Omega) 
    = {}&\frac{\hbar m}{2}\biggl[
        \frac{|D_P(\Omega)|^2}{\Gamma_P\Omega_P}e^{-2r}
        + \Gamma_P\Omega_P\biggl|\frac{D_P(\Omega)}{D_{A\,\rm eff}(\Omega)} - 1\biggr|^2e^{2r} \nonumber \\
        &\hspace{2.5cm}+ 2\frac{|D_P(\Omega)|^2}{|D_{A\,\rm eff}(\Omega)|^2}\tilde{S}_T(\Omega)
      \biggr] , \label{eq:S-F-ser}\\
  S_{\rm par}^F(\Omega) 
    ={}& \frac{\hbar m}{2}\big(
          K_P(\Omega)[K_{A\,\rm{eff}}(\Omega) + 2\tilde{S}_{T}(\Omega)\cosh2r] \nonumber\\
          &\hspace{1cm}+ |D_P(\Omega) - D_{A\,\rm eff}(\Omega)|^2\sinh^2 2r
        \big) \nonumber\\
        &\hspace{1.45cm}\big/\big(K_{A\,\rm{eff}}(\Omega)\cosh2r + 2\tilde{S}_{T}(\Omega)\big) \,. \label{eq:S-F-par}
\end{align}

\subsubsection{Estimates of QBA reduction in GWDs}\label{sec:estimates}

\begin{table*}
  \begin{ruledtabular}
    \begin{tabular}{lll}
      Notation & Quantity & Value used for estimates\\
      \hline
      $J$ & Normalized optical power in the GWD \eqref{J} & $(2\pi\times100\,{\rm Hz})^3$ \\
      $\kappa$ & Interferometer half-bandwidth & $2\pi\times500\,{\rm Hz}$ \\
      $e^{2r}$ & Squeeze factor for the \{serial, parallel\} topology & \{4 (6\,dB), 8 (9\,dB)\} \\
      $\Omega_{A\,\rm eff}$ & Effective resonance frequency of the auxiliary system &
        $-2\pi\times10\,{\rm Hz}$ \\
      $\gamma_A$ & Auxiliary oscillator damping rate in \{mechanical, spin\} implementation & $2\pi\times\{1\,{\rm mHz}$, $3\,{\rm Hz}\}$ \\
      $n_T$ & Thermal occupancy of the \{mechanical, spin\} oscillator & \{2100, 0\}
    \end{tabular}
  \end{ruledtabular}
  \caption{The main parameters and their numerical values used in this paper.}\label{tab:values}
\end{table*}

\begin{figure*}[thb]
\begin{center}
\includegraphics[width=0.95\textwidth]{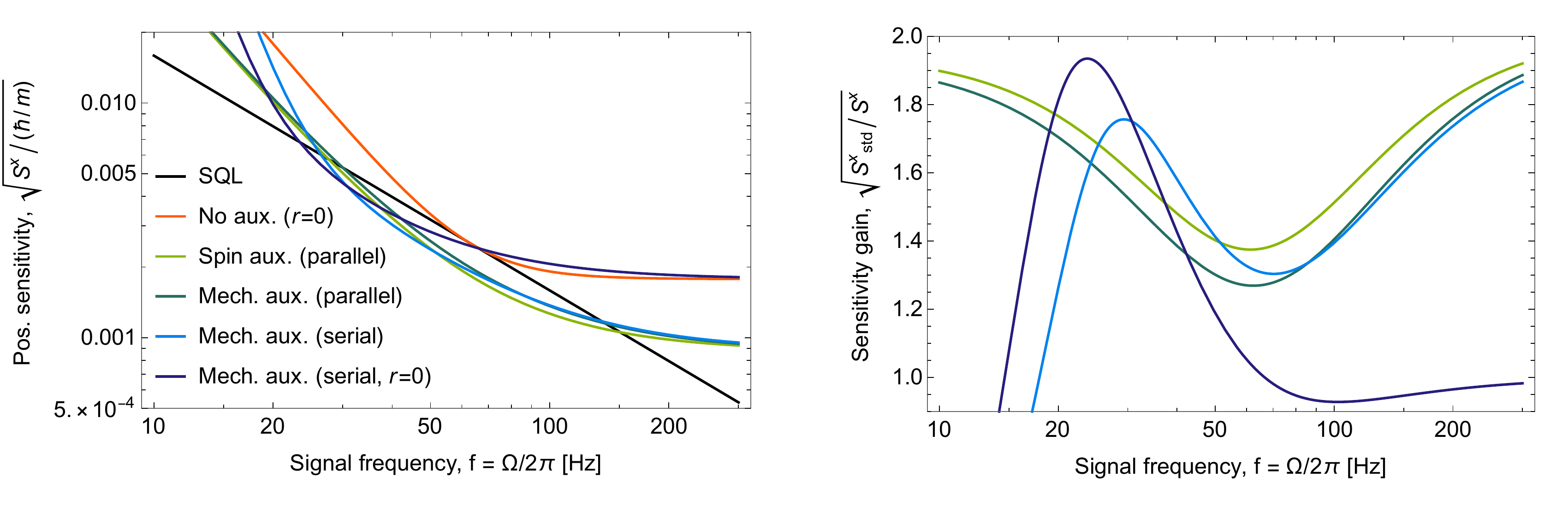}
\caption{Estimates of GWD position sensitivity~\eqref{S^x} improved by quantum noise evasion using an effective auxiliary oscillator with negative mass, see Eqs.~(\ref{eq:S-F-ser}, \ref{eq:S-F-par}). (Left) absolute position sensitivity; (right) sensitivity gain relative to a GWD without a quantum-noise-reducing auxiliary system and without squeezing. The values of the parameters are those listed in Table~\ref{tab:values} unless otherwise noted.}
\label{fig:GWD-sens}
\end{center}
\end{figure*}

We consider here two promising candidate systems for serving as quantum-noise-canceling auxiliary, namely the collective mode of a polarized spin ensemble~\cite{Moeller_Nature_547_191_2017, Thomas_NPhys_17_228_2021} and a high-$Q$ mechanical oscillator~\cite{Mason_NPhys_15_745_2019}. 
The advantage of the spin oscillator is that it can be prepared close to its ground state by optical pumping. The collective spin excitations precess at the Larmor frequency and hence a non-zero oscillator temperature can only be generated by external fields or forces around this frequency. Such forces and fields can be efficiently minimized by magnetic shielding and hence the collective spin temperature close to zero $n_T\approx 0$ can be achieved as demonstrated in Refs.~\cite{Julsgaard_Nature_413_400_2001,Wasilewski_PRL_104_133601_2010,Moeller_Nature_547_191_2017,Thomas_NPhys_17_228_2021}. 
Linewidths of $\gamma_A/2\pi\sim1\text{--}100\,\text{Hz}$ are feasible for spin oscillators~\cite{Wasilewski_PRL_104_133601_2010, Balabas_OE_18_5825_2010}.
In contrast, the thermal occupancy of a mechanical resonator in, e.g., the MHz regime will be of order $n_T\sim10^3$ even for temperatures $T\sim100\,\text{mK}$, whereas very small intrinsic linewidths $\gamma_A/2\pi\sim 1\,\text{mHz}$ are commonplace owing to quality factors $Q\sim10^9$.

The numerical values used here are listed in Table~\ref{tab:values}. We use moderately optimistic values for the spin oscillator damping rate and the thermal occupancy of the mechanical oscillator. The latter corresponds to, for example, a bath temperature of $T=100\,\text{mK}$ and a bare resonance frequency of $\Omega_A/2\pi=1\,\text{MHz}$.

For the sake of generality, we combine the GWD parameters into two effective ones~\cite{12a1DaKh}, the signal-recycled half-bandwidth $\kappa$ and the normalized optical power
\begin{equation}\label{J}
  J = \frac{4\omega_oI_c}{mcL} = \frac{\kappa\Gamma_P\Omega_P}{2} \,,
\end{equation}
where $I_c$ is the total power circulating in both arms of the interferometer, $c$ is the speed of light, and $L$ is the interferometer arms' length. The values of these parameters approximately correspond to the design goal of the Advanced LIGO~\cite{CQG_32_7_074001_2015} and are close to the design values of other advanced GW detectors: Advanced Virgo~\cite{Acernese_CQG_32_024001_2015} and KAGRA~\cite{Aso_PRD_88_043007_2013}. We ignore optical losses for simplicity here; their impact on the sensitivity of the parallel scheme was analyzed in Ref.~\cite{19a1ZePoKh}.

Following the convention used in the GWD community, in our plots we normalize the sum noise spectral densities to the effective displacement signal,
\begin{equation}\label{S^x}
  S^x_{\rm ser,\,par}(\Omega) = \frac{S^F_{\rm ser,\,par}(\Omega)}{m^2\Omega^4} \,.
\end{equation}
The corresponding sensitivity curves achievable using the serial and parallel quantum noise evasion schemes considered in the preceding two subsections are presented in Fig.~\ref{fig:GWD-sens}. They show that the quantum noise evasion techniques allow a significant broadband improvement for signal frequencies of interest $\Omega>\Omega_{\text{low}}$ relative to a ``standard'' interferometer that invokes neither quantum noise evasion nor input squeezing.
These plots also demonstrate the relative merits of the serial and the parallel topologies. 
The degree of squeezing $r$ impacts the performance of the serial and parallel topologies differently; this is because in the parallel case the squeezing is distributed among the two arms out of which the GWD signal enters only one. We choose to compare the two topologies using squeezing levels which render the performance approximately equal in the region dominated by imprecision shot noise; this 
requires squeezing which is stronger by $3\,\text{dB}$ in the parallel case as compared to the serial case (see Ref.~\cite{Ma_NPhys_13_776_2017}).
At the same time, in the serial topology, the auxiliary resonance at $|\Omega_{A\,\rm eff}|/2\pi=10\text{Hz}$ drastically degrades the performance for frequencies $\sim\Omega_{A\,\rm eff}$, while in the parallel topology this effect is suppressed via the optimal combination $\alpha(\Omega)$ of GWD and auxiliary signals in post-processing.

\section{Conclusion and outlook}\label{sec:Conclusion-outlook}

We have presented a framework for engineering QMFSs across subsystems with potentially very different spectral domains by means of periodic modulation of their coupling to light (or another traveling field). Our method applies to disparate oscillators with vastly different resonance frequencies ranging from Hz to GHz (and beyond, in principle), and is applicable to systems which couple to electromagnetic radiation with carrier frequencies ranging from microwaves to the optical domain.

A number of such platforms operating in or near the quantum-coherent regime are emerging at present, such as bulk acoustic wave resonators coupled to superconducting qubits~\cite{Chu_Science_358_199_2017}, optical modes coupled to nanoscale cavities~\cite{, Chang_RMP_90_031002_2018, MacCabe_Science_370_840_2020}, coupled silicon nanobeams~\cite{Beguin_PNAS_117_29422_2020, Fink_QSciTech_5_034011_2020}, and coherent microwave-optical interfaces mediated by nanomechanical transducers~\cite{Mirhosseini_Nature_588_599_2020, Honl2021}.
Our scheme provides the means to combining these systems into a variety of hybrid systems.
Since the QBA-evading measurements (inherent to QMFSs) enable a variety of applications, as detailed in Sec.~\ref{sec:applications}, e.g., sensing beyond the SQL, entanglement generation, and teleportation between remote systems, our generic scheme extends the range of quantum systems in which these applications can be feasibly implemented. 
Considering in particular the outstanding challenge of broadband quantum noise reduction in GWDs, we show how our approach can be used to realize an effective free negative mass as required to form a QMFS with the GWD; this does away with the need for hundred-meters-long narrowband filter cavities~\cite{02a1KiLeMaThVy, wp2020}.

While some discussion of the impact of imperfections, such as thermal oscillator noise and optical losses (including finite detection efficiency), was given in Secs.~\ref{subsec:measurement-scheme} and~\ref{sec:applications}, a full assessment of those is beyond the scope of this work. Such an assessment is best made in the context of a specific implementation and application.

In future work, the approach to forming QMFSs laid out in this work could be explored in more complex quantum network topologies within the framework of Ref.~\cite{Karg_PRA_99_063829_2019}.

\begin{acknowledgments}

The authors thank O.\,Sandberg, Y.\,Chen, and H.\,Miao for reading the manuscript and providing useful remarks. 
This work was supported by the European Research Council Advanced grant QUANTUM-N and by VILLUM FONDEN under a Villum Investigator Grant, grant no.\ 25880. 
The work of F.\,K.\ was supported by the Russian Foundation for Basic Research grant 19-29-11003.

\end{acknowledgments}

\appendix

\section{Oscillator embedded in a cavity}\label{app:eqs_of_motion}

We start with the linearized equations in the rotating-wave approximation (regarding the optical decay) for the standard optomechanical system, which can be found in, e.g., Ref.~\cite{12a1DaKh}:
\begin{subequations}\label{eqs_raw}
  \begin{gather}
    \hat{b}^{c,s}(t)
      = -\hat{a}^{c,s}(t) + \sqrt{2\kappa}\,\hat{q}^{c,s}(t) \,,
        \label{eqs_raw_a} \\
    \fulld{\hat{q}^c(t)}{t} + \kappa\hat{q}^c(t)
      = \sqrt{2\kappa}\,\hat{a}^c(t) \,, \label{eqs_raw_qc} \\
    \fulld{\hat{q}^s(t)}{t} + \kappa\hat{q}^s(t)
      = \sqrt{2\kappa}\,\hat{a}^s(t) + Gq_o(t)\hat{x}(t) \,, \label{eqs_raw_qs} \\
    m\biggl[
        \fulldd{\hat{x}(t)}{t} + 2\gamma\fulld{\hat{x}(t)}{t} + \Omega_0^2\hat{x}(t)
      \biggr]
      = \hbar Gq_o(t)\hat{q}^c(t) + \hat{F}(t) \,.
      \label{eqs_raw_x}
  \end{gather}
\end{subequations}
Here $\hat{a}^{c,s}$, $\hat{b}^{c,s}$, and $\hat{q}^{c,s}$ are the cosine and sine quadratures for the input, output, and intracavity fields, respectively, $\kappa$ is the cavity half-bandwidth, $\hat{x}$ is the oscillator position coordinate, $\hat{F}$ is the sum of all other forces, including the thermal one, $m$ is the mechanical mass (which could be negative), $\gamma$ is the oscillator damping rate, and $G$ is the optomechanical coupling factor. We assume that the optical carrier frequency is equal to the cavity eigenfrequency $\omega_o$, its phase is equal to zero, and that the amplitude of the intracavity pump power,
\begin{equation}
  q_o(t) = \sqrt{2N}k(t) \,,
\end{equation}
varies slowly on the $\omega_o$ timescale; here $N$ is the mean number of quanta in the cavity and $k(t)$ is a dimensionless time-dependent function.

Using the normalized position $\hat{X}$, Eq.~\eqref{norm_XP}, and force $\hat{f}$, Eq.~\eqref{norm_F}, we can recast Eqs.~\eqref{eqs_raw} in the following form:
\begin{subequations}\label{eqs_norm}
  \begin{gather}
    \hat{b}^{c,s}(t) = -\hat{a}^{c,s}(t) + \sqrt{2\kappa}\,\hat{q}^{c,s}(t) \,,
      \label{eqs_norm_a} \\
    \fulld{\hat{q}^c(t)}{t} + \kappa\hat{q}^c(t)
      = \sqrt{2\kappa}\,\hat{a}^c(t) \,, \label{eqs_norm_qc} \\
    \fulld{\hat{q}^s(t)}{t} + \kappa\hat{q}^s(t)
      = \sqrt{2\kappa}\,\hat{a}^s(t) + 2g k(t)\hat{X}(t) \,, \label{eqs_norm_qs} \\
    \frac{1}{\Omega_0}\biggl[
        \fulldd{\hat{X}(t)}{t} + 2\gamma\fulld{\hat{X}(t)}{t} + \Omega_0^2\hat{X}(t)
      \biggr]
      = 2gk(t)\hat{q}^c(t) + \hat{f}(t) \,,\label{eqs_norm_x}
  \end{gather}
\end{subequations}
where we have introduced the pump-enhanced oscillator-cavity coupling rate
\begin{equation}
g\equiv\sqrt{\frac{\hbar}{2\rho}}\sqrt{N}G\,.
\end{equation}

We now assume the bad-cavity regime,
\begin{equation}
  \kappa \gg |\Omega_0|,\,g\max|k(t)| \,.
\end{equation}
In this case, we obtain from Eqs.~\eqref{eqs_norm},
\begin{subequations}\label{eqs_badcav_t}
  \begin{gather}
  \hat{b}^c(t) = \hat{a}^c(t) \,, \\
    \hat{b}^s(t) = \hat{a}^s(t) + \sqrt{\Gamma} k(t)\hat{X}(t) \,, \\
    \frac{1}{\Omega_0}\biggl[
        \fulldd{\hat{X}(t)}{t} + 2\gamma\fulld{\hat{X}(t)}{t} + \Omega_0^2\hat{X}(t)
      \biggr]
      = \sqrt{\Gamma}k(t)\hat{a}^c(t) + \hat{f}(t)  \,, \label{eqs_badcav_t_X}
  \end{gather}
\end{subequations}
where the coupling between the oscillator and the \emph{external} field is parametrized by the rate
\begin{equation}\label{Gamma_def}
  \Gamma = \dfrac{\hbar}{\rho}\dfrac{4NG^2}{\kappa} =\dfrac{8g^2}{\kappa}\,.
\end{equation}
On account of Eq.~\eqref{k_of_t}, the Fourier form of Eqs.~\eqref{eqs_badcav_t} is given by Eq.~\eqref{eq:badcav_Omega_raw}.

\section{Scattering relation for modulated driving}\label{app:effective}

It follows from the assumption~\eqref{RWA2} that $\chi(\Omega)$ and therefore $\hat{X}(\Omega)$ are significant only if $\Omega$ is close to $\pm|\Omega_0|\approx\pm\tilde{\Omega}$. In this case, it follows from Eqs.~\eqref{eq:badcav_Omega_raw} that
\begin{equation}
  \hat{X}(\Omega\pm\tilde{\Omega}) = \chi(\Omega\pm\tilde{\Omega})\biggl[
      \sqrt{\Gamma}\sum_{n=-\infty}^\infty k_{n\pm1}\hat{a}^c(\Omega-n\tilde{\Omega})
      + \hat{f}(\Omega\pm\tilde{\Omega})
    \biggr] ,
\end{equation}
and
\begin{multline}
  \hat{b}^s(\Omega) = \hat{a}^s(\Omega)
    + \sqrt{\Gamma}[k_1\hat{X}(\Omega-\tilde{\Omega}) + k_{-1}\hat{X}(\Omega+\tilde{\Omega})] \\
  = \hat{a}^s(\Omega)
    + \Gamma|k_1|^2[\chi(\Omega-\tilde{\Omega}) + \chi(\Omega+\tilde{\Omega})]\hat{a}^c(\Omega) \\
          + \sqrt{\Gamma}[
                k_1\chi(\Omega-\tilde{\Omega})\hat{f}(\Omega-\tilde{\Omega}) +
                k_{-1}\chi(\Omega+\tilde{\Omega})\hat{f}(\Omega+\tilde{\Omega})
              ] \\
          + \Gamma\sum_{n\ne0}
              [k_1k_{n-1}\chi(\Omega-\tilde{\Omega}) + k_{-1}k_{n+1}\chi(\Omega+\tilde{\Omega})]
              \hat{a}^c(\Omega-n\tilde{\Omega})
         \,.
\end{multline}
Using the approximation \eqref{chi_nb}, we obtain the scattering relation
\begin{multline}
  \hat{b}^s(\Omega) = \hat{a}^s(\Omega)
    + \frac{s\Gamma |k_1|^2\Lambda\hat{a}^c(\Omega)}{\ell^2(\Omega)+\Lambda^2} \\
    + \frac{is\sqrt{\Gamma}}{2}\biggl[
          \frac{k_{-1}\hat{f}(\Omega+\tilde{\Omega})}{\ell(\Omega-\Lambda)}
          - \frac{k_1\hat{f}(\Omega-\tilde{\Omega})}{\ell(\Omega+\Lambda)}
        \biggr]
    + \hat{b}^s_{\rm extra}(\Omega) \,,
\end{multline}
where the term $\hat{b}^s_{\rm extra}(\Omega)$ is given by Eq.~\eqref{bs_extra}. This amounts to Eqs.~\eqref{badcav_Omega_var} and \eqref{eq:eff-params} in the main text as seen by exploiting the relation $\ell(\Omega+\Lambda)\ell(\Omega-\Lambda)=\ell^2(\Omega)+\Lambda^2$.

\section{Evolution vs.\ resonance frequency and their compensation}\label{app:chi_eff}

\subsection{Evolution vs.\ resonance frequency in the effective susceptibility~\eqref{eq:chi-eff_raw}}

The effective susceptibility~\eqref{eq:chi-eff_raw} contains two, generally distinct frequencies, the evolution frequency $\Omega_{\text{eff}}$ in the numerator, and the resonance frequency $\sqrt{\Omega_{\text{eff}}^2+\gamma^2}$ in the denominator.

The evolution frequency $\Omega_{\text{eff}}$ in the numerator of \eqref{eq:chi-eff_raw} represents the fact that QBA acts on the oscillator variable conjugate to the observable, and hence is only observed due to the dynamical rotation at angular frequency $\Omega_{\text{eff}}$ of the canonical oscillator position and momentum into each other; this points to the fact that the limit of a single-quadrature measurement is $\Omega_{\text{eff}}\rightarrow 0$.
In this case the scattering relation~\eqref{badcav_Omega_var} reduces since the nominal QBA term $\propto\chi_{\text{eff}}\hat{a}^c(\Omega)$ vanishes whereas the thermal force readout $\propto\chi_{\text{eff}}\hat{f}_{\text{eff}}$ is finite. This is to expected since in the particular case $\Omega_{\text{eff}}=0\Leftrightarrow\Lambda=0$ our (generally) ``detuned'' periodic driving scheme reduces to the ``resonant'' class of coupling envelopes,
of which familiar examples are the stroboscopic and two-tone measurements of a \emph{single} oscillator quadrature discussed in the Introduction.
In this case all the (nominal) QBA goes to the unobserved oscillator quadrature, accompanied by the conditional squeezing of the observed quadrature [which amounts to a QBA-free measurement of a single phase of the force $\sin(\Omega_0 t-\Phi)\hat{f}(t)$ as follows from Eq.~\eqref{eq:f-eff}]. Naturally, such single-quadrature measurements are not limited by the Heisenberg uncertainty relation; in some sense this can be viewed as a degenerate, 1-dimensional example of a QMFS, whereas we consider the 2-dimensional example captured by Eqs.~\eqref{eq:EPR-osc} and~\eqref{eq:EPR-comm} to belong to the class of non-degenerate lowest-dimensional cases.

Turning now to the resonance frequency of $\chi_{\text{eff}}$, $\sqrt{\Omega_{\text{eff}}^2+\gamma^2}$, we note that it contains a shift from the intrinsic decay rate $\gamma$ due to the fact that both \emph{effective} oscillator variables $\hat{\mathcal{X}}$ and $\hat{\mathcal{P}}$~[Eq.~\eqref{eq:slowly-rot-vars}] experience decay, whereas for the bare oscillator, viscous damping $\dot{\hat{P}}=-2\gamma\hat{P}+\cdots$ alone was assumed, which led to the susceptibility in Eq.~\eqref{chi_norm}.

\subsection{Compensation of the resonance frequency by parametric excitation}

Here we show that, using the parametric excitation, it is possible to effectively redistribute the damping between $\hat{\mathcal{X}}$ and $\hat{\mathcal{P}}$ (keeping the total damping unchanged) and, in particular, concentrate it in the ``$\hat{\mathcal{X}}$'' channel. To this end, consider again the equation of motion~\eqref{eqs_badcav_t}. We assume for simplicity the optimal two-tone shape of the coupling envelope $k(t)=\sqrt{2}\cos\tilde{\Omega}t$, and suppose that the resonance frequency $\Omega_0$ is modulated at the frequency $2\tilde{\Omega}$,
\begin{multline}
  \fulldd{\hat{X}(t)}{t} + 2\gamma\fulld{\hat{X}(t)}{t}
    + \Omega_0^2\left(1 + \frac{4\mu}{\Omega_0} \sin2\tilde{\Omega}t\right)\hat{X}(t) \\
    = \Omega_0[\sqrt{2\Gamma}\hat{a}^c(t)\cos\tilde{\Omega}t + \hat{f}(t)] \,,\label{eq_X_param}
\end{multline}
where $\mu$ is the rescaled parametric modulation factor.
Introduce the rotating-frame amplitudes with respect to $\tilde{\Omega}$ as follows [note that this frame is different from that defined by Eq.~\eqref{eq:slowly-rot-vars}]:
\begin{subequations}
  \begin{align}
    \hat{X}(t) &= \hat{\mathcal{X}}(t)\cos\tilde{\Omega}t + \hat{\mathcal{P}}(t)\sin\tilde{\Omega}t \,, \\
    \fulld{\hat{X}(t)}{t}
      &= \tilde{\Omega}[-\hat{\mathcal{X}}(t)\sin\tilde{\Omega}t + \hat{\mathcal{P}}(t)\cos\tilde{\Omega}t]\,, \\
    \hat{f}(t) &= \hat{f}^c(t)\cos\tilde{\Omega}t + \hat{f}^s(t)\sin\tilde{\Omega}t \,.
  \end{align}
\end{subequations}
Substitute them into Eq.~\eqref{eq_X_param} and neglect the fast-oscillating terms to get
\begin{subequations}\label{shortened_eqs}
  \begin{align}
    \fulld{\hat{\mathcal{X}}(t)}{t} + (\gamma-\mu)\hat{\mathcal{X}}(t)
      - \Lambda\hat{\mathcal{P}}(t) &= -\frac{f^s(t)}{2} \,, \\
    \fulld{\hat{\mathcal{P}}(t)}{t} + (\gamma+\mu)\hat{\mathcal{P}}(t)
      + \Lambda\hat{\mathcal{X}}(t) &= \sqrt{\frac{\Gamma}{2}}\hat{a}^c(t)
      + \frac{f^c(t)}{2} \,.
  \end{align}
\end{subequations}
Combining these equations and rewriting the result in the Fourier picture, we obtain
\begin{equation}
  \hat{\mathcal{X}}(\Omega) = \chi_{\rm eff}(\Omega)
    \biggl[\sqrt{\frac{\Gamma}{2}}\hat{a}^c(\Omega) + \hat{f}_{\rm eff}(\Omega)\biggr] \,,
\end{equation}
where
\begin{equation}\label{chi_eff_param}
  \chi_{\rm eff}(\Omega) = \frac{\Omega_{\rm eff}}
    {\Omega_{\rm eff}^2 + \gamma^2 - \mu^2 - \Omega^2 - 2i\Omega\gamma}
\end{equation}
is the effective susceptibility and
\begin{equation}
  \hat{f}_{\rm eff}(\Omega)
  = \frac{\Lambda\hat{f}^c(\Omega) - (-i\Omega + \gamma + \mu)\hat{f}^s(\Omega)}{2\Lambda}
\end{equation}
is the effective thermal noise with the spectral density
\begin{equation}\label{S_T_eff_param}
  S_{T\,{\rm eff}}(\Omega)
  = \frac{ \Omega_{{\rm eff}}^2 + (\gamma+\mu)^2 + \Omega^2}{2\Omega_{\text{eff}}^2}S_T \,;
\end{equation}
compare with Eqs.~(\ref{badcav_Omega_var}, \ref{eq:chi-eff_raw}, \ref{S_T_eff_raw}).
Making the particular choice of $\mu=-\gamma$, Eqs.~(\ref{chi_eff_param}, \ref{S_T_eff_param}) yield Eqs.~(\ref{eq:chi-eff}, \ref{S_T_eff}).

\bibliography{abbots_my,biblio_us,khalili_u,ligo,mqm,misc_u}

\end{document}